\documentclass[12pt,preprint]{aastex}
\usepackage{graphicx}
\usepackage{epstopdf}

\slugcomment{Accepted in AJ (Oct 31, 2010)}

\begin{document}

\title{Millimeter and submillimeter high angular resolution\\ interferometric
observations: dust in the heart of IRAS~18162-2048}

\author{M. Fern\'andez-L\'opez\altaffilmark{1}}
\author{S. Curiel\altaffilmark{1}}
\author{J.M. Girart\altaffilmark{2}}
\author{P.T.P. Ho\altaffilmark{3,4}}
\author{N. Patel\altaffilmark{4}}
\and
\author{Y. G\'omez\altaffilmark{5}}

\altaffiltext{1}{Instituto de Astronom{\'\i}a, Universidad Nacional Aut\'onoma de M\'exico
(UNAM), Apartado Postal 70-264, 04510 M\'exico, DF, M\'exico;
manferna@gmail.com, scuriel@astroscu.unam.mx}
\altaffiltext{2}{Institut de Ciencies de l'Espai, (CSIC-IEEC),Campus UAB, Facultat de Ciencies, Torre C5-parell 2, 08193 Bellaterra, Catalunya, Spain; girart@ieec.cat}
\altaffiltext{3}{Academia Sinica Institute of Astronomy and Astrophysics, P.O. Box 23-141, Taipei 10617, Taiwan}
\altaffiltext{4}{Harvard-Smithsonian Center for Astrophysics, 60 Garden Street, Cambridge, MA 02138, USA}
\altaffiltext{5}{Centro de Radioastronom\'{\i}a y Astrof\'{\i}sica, UNAM,
Apartado Postal 3-72, Morelia, Michoac\'an 58089, M\'exico;
y.gomez@astrosmo.unam.mx}

\begin{abstract}
The GGD27 complex includes the HH 80-81-80N system, which is one of the most powerful molecular outflows associated with a high mass star-forming region observed up to now. This outflow is powered by the star associated with the source IRAS 18162-2048. Here we report the detection of continuum emission at sub-arcsec/arcsec resolution with the Submillimeter Array at 1.36~mm and 456~$\mu$m, respectively. We detected dust emission arising from two compact cores, MM1 and MM2, separated by about 7$\arcsec$ ($\sim$12000~AU in projected distance). MM1 spatially coincides with the powerful thermal radio continuum jet that powers the very extended molecular outflow, while MM2 is associated with the protostar that drives the compact molecular outflow recently found in this region. 

High angular resolution obervations at 1.36~mm show that MM1 is unresolved and that MM2 splits into two subcomponents separated by $\sim 1\arcsec$. The mass of MM1 is about 4~M$_{\sun}$ and it has a size of $\lesssim300$~AU. This is consistent with MM1 being associated with a massive and dense (n${\rm (H_2)} \ga 10^{9}$~cm$^{-3}$) circumstellar dusty disk surrounding a high-mass protostar, which has not developed yet a compact HII region. On the other hand, the masses of the two separate components of MM2 are about 2~M$_{\sun}$ each. One of these components is a compact core with an intermediate-mass young protostar inside and the other component is probably a pre-stellar core.

MM1 is the brigthest source at 1.36~mm, while MM2 dominates the emission at 456~$\mu$m. These are the only (sub)millimeter sources detected in the SMA observations. Hence, it seems that both sources may contribute significantly to the bolometric luminosity of the region. Finally, we argue that the characteristics of these two sources indicate that MM2 is probably in an earlier evolutionary stage than MM1.
\end{abstract}


\section{INTRODUCTION}
\medskip

It is well known that low-mass stars form by accretion via circumstellar disks. In addition, collimated outflows are believed to remove excess angular momentum from the system. However, the formation process of high-mass stars ($>8M_{\sun}$) is yet unclear (see \markcite{2007ARA&A..45..481Z}{Zinnecker} \& {Yorke} 2007, for a recent review). When a protostar accretes enough mass to ignite hydrogen, it starts emitting a large amount of ionizing photons. This radiation ionizes the surrounding gas thereby forming an H~II region. At first glance, the pressure from the stellar radiation and the ionized gas, could prevent further accretion (see e.g., \markcite{1971A&A....13..190L,1974A&A....37..149K}{Larson} \& {Starrfield} 1971; {Kahn} 1974). However, accretion through disks may circumvent these problems, allowing protostars to grow up to 60-100~M$_{\sun}$ (\markcite{1989ApJ...345..464N,1996ApJ...462..874J,1999ApJ...525..330Y}{Nakano} 1989; {Jijina} \& {Adams} 1996; {Yorke} \& {Bodenheimer} 1999). 
Several theoretical models have been proposed to explain the formation of high-mass stars (see \markcite{2007ARA&A..45..481Z}{Zinnecker} \& {Yorke} 2007 for a review). Models based on accretion require high accretion rates (e.g., \markcite{2003ApJ...585..850M}{McKee} \& {Tan} 2003), and/or accretion of ionized material after the hydrogen burning has started in the star (\markcite{2006ApJ...637..850K}{Keto} \& {Wood} 2006). More dynamical models are based on the interaction between the collapsing fragments of the parental cloud (i.e., competitive accretion, \markcite{2001MNRAS.323..785B}{Bonnell} {et~al.} 2001), or coalescence of less massive protostars (\markcite{1998MNRAS.298...93B}{Bonnell}, {Bate}, \&  {Zinnecker} 1998). At present, the observational data do not allow to distinguish between these models.

Because of their short lifetimes, high-mass protostars remain embedded in their natal envelopes. This situation hampers the detection and characterization of disks, since disk emission is usually mixed with that of its own envelope. In addition, high-mass protostars are located typically a few kiloparsec away and they usually form in clusters, which makes it difficult to resolve without sufficient angular resolution. Hence, currently there are only a handful of appropriate candidates for detailed studies.

There is some observational evidence of disks around B-type protostars (with stellar masses $\lesssim20$~M$_{\sun}$, see e.g., \markcite{2007prpl.conf..197C,2005IAUS..227..135Z}{Cesaroni} {et~al.} 2007; {Zhang} 2005). Disks around O-type protostars, are elusive. Instead of disks, O-type protostars (with masses $\gtrsim20$~M$_{\sun}$) are surrounded by $\sim10000$~AU rotating molecular structures, which seem to be dynamically unstable or collapsing structures (e.g., \markcite{2006Natur.443..427B,2009ApJ...698.1422Z}{Beltr{\'a}n} {et~al.} 2006; {Zapata} {et~al.} 2009).

In the Sagittarius region, at a distance of $1.7$~kpc (\markcite{1980ApJ...235..845R}{Rodr\'{\i}guez} {et~al.} 1980), one of the most powerful radio continuum jets has been associated with high mass star formation. The HH~80-81-80N system, as part of the GGD27 complex, has several remarkable characteristics:
(i) the length of the thermal jet, $\sim$5.3 pc, is one of the largest found so far (\markcite{1993ApJ...416..208M}{Mart\'{\i}}, {Rodr\'{\i}guez}, \&  {Reipurth} 1993). Multiwavelength VLA studies (\markcite{1993ApJ...416..208M,1995ApJ...449..184M,1998ApJ...502..337M}{Mart\'{\i}} {et~al.} 1993; {Mart\'{\i}}, {Rodr\'{\i}guez}, \&  {Reipurth} 1995, 1998, 1999) show that this radio jet ends at two Herbig-Haro objects to the south (HH80 and 81; \markcite{1988A&A...202..219R}{Reipurth} \& {Graham} 1988) and at a radio source to the north (HH 80 North; \markcite{1993ApJ...416..208M,1994ApJ...435L.145G,2001ApJ...562L..91G,2009ApJ...695.1505M}{Mart\'{\i}} {et~al.} 1993; {Girart} {et~al.} 1994, 2001; {Masqu{\'e}} {et~al.} 2009). 
(ii) The HH 80-81 objects are very bright in the optical (\markcite{1998AJ....116.1940H}{Heathcote}, {Reipurth}, \&  {Raga} 1998) and radio wavelengths \markcite{1989RMxAA..17...59R}({Rodr{\'{\i}}guez} \& {Reipurth} 1989), $\sim10$ times brighter than those of classical HH objects \markcite{1995RMxAC...1...59C}({Curiel} 1995). 
(iii) The motions found in these HH objects and related ejections are very large (optical line widths up to $700$ km~s$^{-1}$, \markcite{1988A&A...202..219R}{Reipurth} \& {Graham} 1988; proper motions up to $\sim$500 km~s$^{-1}$, \markcite{1998ApJ...502..337M}{Mart\'{\i}} {et~al.} 1998). 

This radio jet may be driving a molecular outflow detected by single dish CO observations (\markcite{1989ApJ...347..894Y}{Yamashita} {et~al.} 1989; \markcite{2001A&A...378..495R}{Ridge} \& {Moore} 2001; \markcite{2004MNRAS.347..295B}{Benedettini} {et~al.} 2004). The northern lobe of the outflow approaches us (blue lobe) while the southern one moves away from us (red lobe). Recent Spitzer observations show a V-shaped biconical structure at 8~$\mu$m roughly coincident with the center of the molecular outflow (\markcite{2008ApJ...685.1005Q}{Qiu} {et~al.} 2008). The vertex of this biconical structure coincides with the radio jet.

The powering source of the radio jet and the CO outflow, coincides with the IRAS source 18162-2048, which has been observed at 20 cm (\markcite{1993ApJ...416..208M}{Mart\'{\i}} {et~al.} 1993) and at 5.8~$\mu$m (\markcite{2008ApJ...685.1005Q}{Qiu} {et~al.} 2008). Based on its IRAS fluxes, the bolometric luminosity of this source is $\sim1.9\times 10^4$~L$_{\sun}$, corresponding to a B0 zero age main sequence (ZAMS) star. Other typical tracers of young massive stars have been found in this region, such as compact CS emission (\markcite{1991ApJ...373..560Y}{Yamashita} {et~al.} 1991) and  bright H$_2$O and CH$_3$OH masers (\markcite{1978ApJ...226..115R}{Rodr\'{\i}guez} {et~al.} 1978, 1980; \markcite{1995ApJ...453..268G,2000MNRAS.317..315V,2004ApJS..155..149K}{G{\'o}mez}, {Rodr\'{\i}guez}, \&  {Mart\'{\i}} 1995; {Val'tts} {et~al.} 2000; {Kurtz}, {Hofner}, \&  {{\'A}lvarez} 2004). Ground-based infrared observations have identified this source as part of a cluster of intermediate-to-high mass young stars (\markcite{1992A&A...266..219A,1994A&A...292L...9A,1997ApJ...479..339S}{Aspin} \& {Geballe} 1992; {Aspin} {et~al.} 1994; {Stecklum} {et~al.} 1997) and at the same time as the responsible for illuminating the reflection nebula seen at these wavelengths (\markcite{1991A&A...252..299A,1992A&A...266..219A}{Aspin} {et~al.} 1991; {Aspin} \& {Geballe} 1992).

BIMA observations at 1.4 and 3.5~mm show compact dust emission ($<3\arcsec$ in size), as well as compact SO($5_{5}-4_{4}$) line emission ($<5\arcsec$ in size) toward the driving source of the outflow (\markcite{2003ApJ...597..414G}{G{\'o}mez} {et~al.} 2003). These observations also show that there is a second millimeter source about $7\arcsec$ to the NE from the center of the radio jet, with weak continuum emission at 1.4~mm, coinciding with warm NH$_3$(1,1) and HCN(1-0) emission. This source nearly coincides with a weak radio continuum compact source detected at 3.5 cm and a strong water maser spot (named VLA3 in \markcite{1995ApJ...453..268G}{G{\'o}mez} {et~al.} 1995). It is also associated with a very young one-side molecular outflow observed at high velocity towards the south-east (\markcite{2009ApJ...702L..66Q}{Qiu} \& {Zhang} 2009).

In this paper, we report Submillimeter Array\footnote{The Submillimeter Array is a joint project between the Smithsonian Astrophysical Observatory and the Academia Sinica Institute of Astronomy and Astrophysics and is funded by the Smithsonian Institution and the Academia Sinica.} (SMA; \markcite{2004ApJ...616L...1H}{Ho}, {Moran}, \& {Lo} 2004) observations of the GGD27 system at 1.36~mm and 456~$\mu$m. 
Pushing at longer wavelengths allows the study of colder (10-100~K) regions. The SMA has the ability of observing at two frequencies simultaneously, sampling the (sub)millimeter spectrum of the sources with a single experiment. The SMA also allows higher angular resolution ($<1\arcsec$), which is crucial for defining structures and resolving the multi-component systems. The SMA also has a very wide spectral band and can detect many spectral lines simultaneously.
We have applied different phase calibration techniques to the submillimeter data. We also obtained spectral-line data which we will publish elsewhere. The observations are described in \S 2 and the results are presented in \S 3. A discussion on the analyzed data is presented in \S 4 and finally, a summary with the main conclusions is presented in \S 5.

\section{OBSERVATIONS} 

The observations were conducted in two epochs separated by two years (2005-2007). The details of the observations are shown in Table \ref{tab: T1}. During the first epoch we observed with the SMA in its compact configuration while using the dual frequency mode (220/658~GHz). During the second epoch we observed only at 220~GHz with the SMA in its very extended configuration. The phase center was RA(J2000.0)$= 18^h 19^m 12\fs099$ and DEC(J2000.0)$= -20\degr 47\arcmin 30\farcs00$, near the position of the source at the center of the thermal radio continuum jet of the GGD27 system.

On 2005 August 24, the two SMA receiver bands were tuned simultaneously at 215/225~GHz and 648/658~GHz (each receiver band has two sidebands with a bandwidth of 1.968~GHz each, and separated by 10~GHz; they are labeled lower and upper sideband respectively, LSB/USB). The interferometer was in its compact configuration and the projected baselines ranged from 13 to 68~m. This configuration produced a synthesized beam size of $8\farcs1\times3\farcs0$ at 220~GHz and of $2\farcs4\times1\farcs1$ at 658~GHz. The SMA primary beam has a size of about $50\arcsec$ and 17$\arcsec$ at 220~GHz and 658~GHz, respectively. Weather was good during the observations, with a relative humidity of $7\%$ and a zenith opacity ranging between $\tau _{225}=0.05$ and 0.08 ($\tau _{690}\sim$0.9 - 1.5, following \markcite{1994ASPC...59...87M}{Masson} 1994), measured with the radiometer of the nearby Caltech Submillimeter Observatory, at 225~GHz. The recorded data were edited, calibrated, imaged and analyzed using the Miriad (\markcite{1995ASPC...77..433S}{Sault}, {Teuben}, \&  {Wright} 1995) and AIPS (developed by NRAO) packages.

On 2007 May 29, we observed with the low frequency receiver tuned at the same frequency as in the previous observation (215/225~GHz). The array was in its very extended configuration with projected baselines ranging between 50 and 514~m, yielding an angular resolution of $0\farcs7\times0\farcs4$. At this time, the zenith opacity during the observations was $\tau _{225}=0.19$ on average. This second set of data was also edited, calibrated, imaged and analyzed using the Miriad and AIPS packages.

To image the data we applied a u-v weighting during the Fourier transformation stage. The robust parameter (\markcite{1995_Briggs,1999_Briggs}{Briggs} 1995; {Briggs}, {Schwab}, \& {Sramek} 1999) controls this weighting, relating the weight of a u-v cell to the number of visibilities within it. The robust parameter ranges between -2 and 2 in Miriad. A robust of 2 corresponds to a natural weighting and a robust of -2 corresponds to a uniform weighting.  While the natural weighting gives a lower rms noise level, the uniform weighting minimizes the sidelobe levels and gives a better spatial resolution. Values of the Miriad's robust parameter between -2 and 2 provide a compromise between these two extrema.

\subsection{SMA $220$~GHz observations}
\subsubsection{Compact configuration}
At 220~GHz the system temperature was between 70 and 150~K, mainly depending on the elevation of each source along the $\sim$9 hours observing track (about 3 hours of which were on target source). The quasar 3C454.3 was used to correct the bandpass and Callisto was used to do the absolute flux calibration. The observations were carried out in cycles of 25 minutes, observing VX Sgr for 10 minutes (in order to try a phase-transfer calibration), then moving to GGD27 for 10 minutes and finally observing the quasar J1911-201 for 5 minutes in each cycle. The gain solutions obtained for the quasar J1911-201 (with flux densities of 2.13 and 2.18~Jy at 215 and 225~GHz, respectively) were applied to GGD27 and VX Sgr. The flux calibration was performed with the visibility amplitudes from Callisto and its uncertainty was estimated to be less than 20\% by comparing measured and expected fluxes for several observed planets and moons (Ganymede, Neptune, Ceres and Callisto; see Table \ref{tab: T9}). Dirty maps were obtained after discarding one of the seven initial antennas because it was unable to follow the atmospheric-phase behavior properly (probably due to instrumental problems). The channels with line emission were removed from the bandwidth before making a continuum channel for each sideband. Then, both sidebands (with similar rms noise levels) were combined and self calibration applied with an averaging time of 3 minutes. An image of the region was then cleaned and restored, applying natural weighting. The resultant image has a synthesized beam size of $8\farcs1\times 3\farcs0$ (P.A.$=34\fdg1$) and an rms of $\sim10$ mJy beam$^{-1}$, which appears to be limited by the dynamic range. The astrometric accuracy was $0\farcs1$. In addition, nine molecular lines (five SO$_2$ transitions, two lines of SO, one of C$^{17}$O and another of H$_2$CO) were detected through both sidebands. The results of the analysis of these lines will be published in a subsequent paper.
\subsubsection{Very extended configuration}
During the 2007 May 29 observations, the system temperature ranged between 150 and 400~K. 3C454.3 was used for bandpass calibration. The frequency setup was identical to the compact configuration and seven of the SMA antennas were used on this occasion. The phase and amplitude calibrator was also J1911-201, with a measured flux density of 1.24~Jy at 215~GHz and 1.16~Jy at 225~GHz. The fluxes at 225~GHz of 3C454.3 and MWC349 (which was also included as a flux calibrator) were 8.57 and 1.97 Jy respectively. Thus, comparing these values with those found for these calibrators in the SMA data base, the uncertainty in the flux scale was estimated to be less than 15\%. After the visibilities were calibrated and channels with line emission were removed, dirty maps of the continuum emission were constructed. Then the two sidebands were combined and self calibrated with an averaging time of 3 minutes. An image was cleaned and restored applying natural weighting. The synthesized beam of the resulting image has a size of $0\farcs68\times 0\farcs38$ (P.A.$=13\fdg0$) and the rms of the image is $\sim$3.0 mJy beam$^{-1}$. The accuracy on the position of the sources based on phase noise was $\sim0\farcs02$, which coincides with the absolute positional accuracies of the calibrators observed. At least seven of the nine molecular lines observed in the compact configuration were also detected at this epoch.
\subsection{SMA 658~GHz observation}
On 2005 August 24 we observed simultaneously with two receiver bands tuned at 215/225~GHz and 648/658~GHz. At 658~GHz, the system temperature varied between 2000 and 4000~K, more than one order of magnitude higher than that at 220~GHz due to the higher atmospheric emission at 658~GHz. The bandpass was calibrated finding a solution for each baseline of the quasar 3C454.3 (with the Miriad task UVGAINS), which had a flux density of 13.5~Jy at this frequency, while the amplitude gains were obtained with visibility data from Ceres, which had the best signal-to-noise ratio among all calibrators observed at this frequency. The estimated flux uncertainty is less than 15\%. Astrometric accuracy is estimated to be $0\farcs2$. Given that interferometric observations at high frequencies, such as 650~GHz, are still very difficult, observations were planned to be able to phase-calibrate the data: a) using the standard procedure (using the quasar J1911-201 as the phase calibrator), b) using the water maser at 658 GHz associated with a nearby evolved star (VX
Sgr) to phase-calibrate the data, and c) applying a phase transfer from the simultaneous observations at 220 GHz.
\subsubsection{Standard calibration}
One of the main problems concerning submillimeter calibrations is the lack of strong calibrators at high frequencies. At the time of observations, the quasar J1911-201, used without problems as phase calibrator at 220~GHz, was too weak ($<1.5$~Jy) to be detected at 658~GHz. Thus, the standard procedure of phase calibration at high frequency was not possible.
\subsubsection{Calibration using a close by H$_2$O maser}
The strong 658.007~GHz H$_2$O 1$_{1,0}$-1$_{0,1}$ maser line of VX Sgr (a red supergiant) was used as phase calibrator for the USB high frequency sideband. VX Sgr is $\sim$3$\degr$ away from GGD27 and this fact together with the fairly good weather, made the calibration cycle time of 25 minutes good enough to remove the slow instrumental and atmospheric drifts. To find the best gain solutions, only the strongest channels with a stable phase were used (averaging the line channels into a continuum channel). The visibilities from 6 antennas were used to self calibrate the water maser (one of the antennas did not follow the phase calibration properly, probably due to instrumental problems). Then, the phase-solutions were applied to the GGD27 upper sideband data. A cleaned image of the upper sideband shows emission consistent with the sources detected at 220~GHz. Then, the continuum emission was self calibrated to correct for the rapid tropospheric fluctuations, improving the signal-to-noise ratio (SNR) by a factor of $\sim4$. No spectral lines were detected at this frequency. The image was natural weighted, giving a synthesized beam of $2\farcs41\times 1\farcs14$ with a P.A.$=37\fdg1$ (see Fig. \ref{fig: 650g}). 

Two compact continuum sources were clearly detected at this frequency. In order to properly determine the flux of the sources (see Table \ref{tab: T2}), a correction for the attenuation of the primary beam was applied to the final image. The rms (measured into the primary beam and out of the sources) before and after the primary beam correction was 457 and 678 mJy beam$^{-1}$.
\subsubsection{Phase transfer from low frequencies}
A different way to do phase calibration was tried by transferring the phase solutions from 215 to 658~GHz. This method requires very good quality data and the stability of the response of the detectors. The phase calibration was tried using two strong masers of VX Sgr (the water maser at 658~GHz and a SiO maser at 215~GHz), but a high rms noise level of the phase at high frequency on GGD 27, prevented the success of the calibration of this source. However, we report the success of this calibration technique on the strong water maser of VX Sgr. The details of this calibration are presented in the appendix.
\subsubsection{Positional corrections}
Comparing the continuum images of GGD27, obtained at 220 and 658~GHz, we found that the peak positions of the two observed sources in the high frequency image were offset about $0\farcs7$ to the NW with respect to their positions found in the low frequency image. In addition, the coordinates of the VX Sgr evolved star and its phase center did not match, showing a similar trend as the offset in the GGD27 image. Since we used the water maser at 658~GHz associated with VX Sgr for the phase calibration of GGD27, an artificial offset was transferred to the GGD27 image.

In order to correct this positional offset, we obtained the peak position of the centroid of the 215~GHz SiO maser of VX Sgr and compared it with the coordinates of the phase center used to observe this evolved star. This comparison indicates that the SiO maser is located about $0\farcs80$ to the north and $0\farcs01$ to the west from the coordinates used during the observations. A similar shift is found when comparing the coordinates of this star from SIMBAD database and the coordinates of the phase center for VX Sgr ($0\farcs69$ to the north and $0\farcs01$ to the west). Since the spatial distribution of the SiO and H$_2$O maser spots is small ($\lesssim0\farcs4$) compared with the synthesized beam size of the observations, we assume that the centroid of both types of masers coincide spatially. We applied the estimated positional offset of the centroid of the SiO maser to the calibrated water maser position at 658~GHz, obtaining that the new position of the H$_2$O maser coincides with the SiO maser position within the pixel size (0\farcs18) at 658~GHz. Then, we applied the same correction to the GGD27 data. The coordinates of the two submillimeter sources in the corrected high frequency image spatially coincide with those obtained at 1.36~mm, within $0\farcs2$, the astrometric accuracy of the 658~GHz image. The high frequency image presented here (see Fig. \ref{fig: 650g}) was corrected as described above. 

\subsection{Notes on flux calibration}
Table \ref{tab: T9} shows the measured and expected flux densities for all the calibrators used during these observations. We find that in all epochs, the obtained flux densities coincide with those expected within a 15\% error. The exceptions are Ceres for the 2005 August observations at 220~GHz, and Neptune for the 2005 August observations at 658~GHz (the latter was partially resolved by the interferometer). For the May 2007 observations towards 3C454.3 (the strongest calibrator) we obtain an exceptional 3.8\% accuracy at 220~GHz, compared to the interpolated data measured between 215 and 225~GHz from the SMA quasar monitoring. Finally, positional accuracy of the calibrators is better than $0\farcs1$.

\section{RESULTS AND ANALYSIS}
\subsection{Continuum emission}
\begin{figure}[h]
\centering
\includegraphics[width=0.42\textwidth, angle=270]{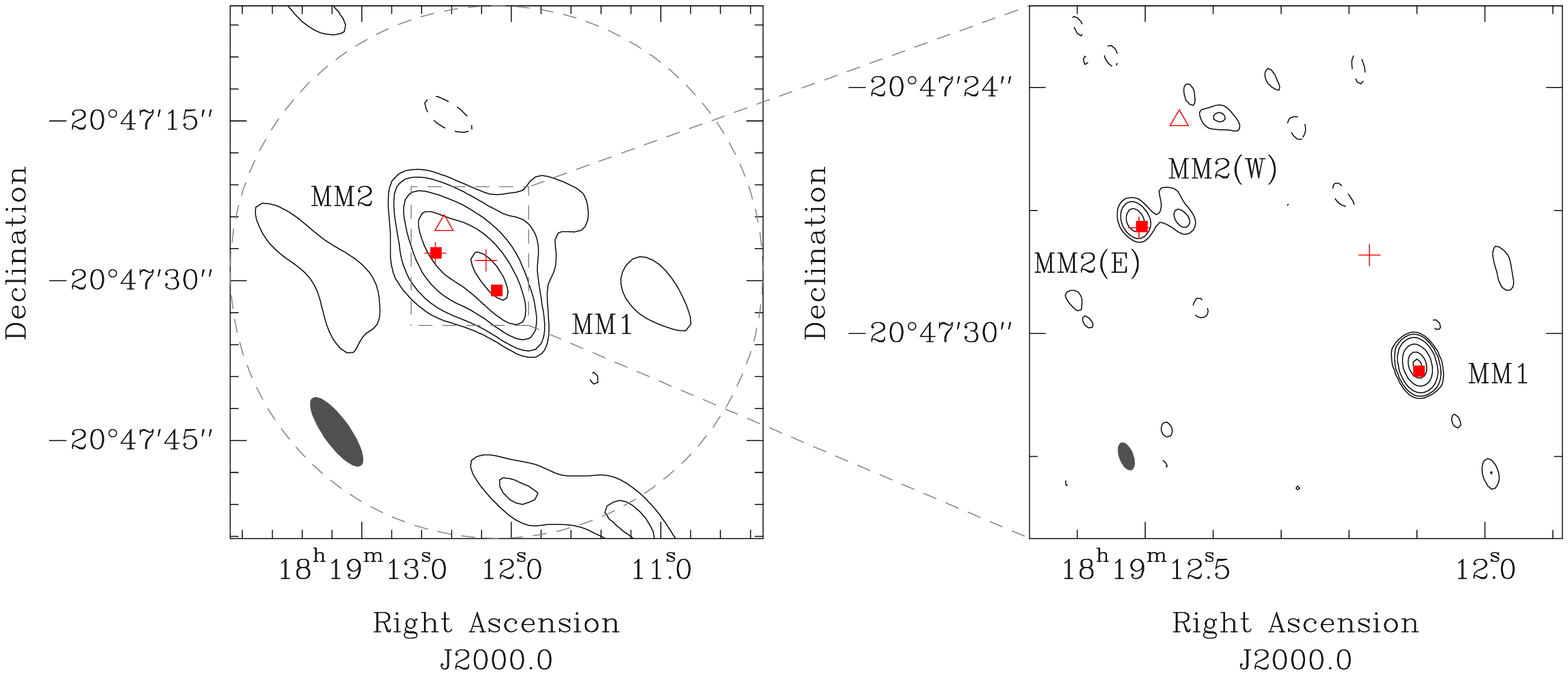}
\caption{1.36~mm (220~GHz) continuum images of the region. The left panel shows the low angular resolution image and the right panel shows the high angular resolution image, both obtained with natural weighting. Contours are -5, -3, 3, 5, 10, 20 and 40 $\times 10.0$~mJy beam$^{-1}$, the rms noise of the low angular resolution data and -5, -3, 3, 5, 10, 30, 75 and 120 $\times3.0$~mJy beam$^{-1}$, the rms noise of the high angular resolution data, respectively. The filled squares mark the positions of the sources detected at 3.5~cm (\markcite{1995ApJ...453..268G}{G{\'o}mez} {et~al.} 1995), the crosses mark two water masers (\markcite{1995ApJ...453..268G}{G{\'o}mez} {et~al.} 1995; \markcite{2005AJ....130..711K}{Kurtz} \& {Hofner} 2005) and the triangle marks the position of a CH$_3$OH class I maser (\markcite{2004ApJS..155..149K}{Kurtz} {et~al.} 2004). The positional accuracy of these sources is $0\farcs1$, except for the water maser located to the north of MM1, on the radio jet tip at 3.5~cm (see Fig. \ref{fig: 650g}), which is $0\farcs01$ (\markcite{2005AJ....130..711K}{Kurtz} \& {Hofner} 2005). The synthesized beam are shown as a gray solid ellipse at the bottom left corner of each image. The dashed circle shows the full width half maximum contour of the primary beam (which has a radius of $25\arcsec$).
}
\label{fig: 220g}
\end{figure}

\begin{figure}[h]
\centering
\includegraphics[width=0.9\textwidth, angle=270]{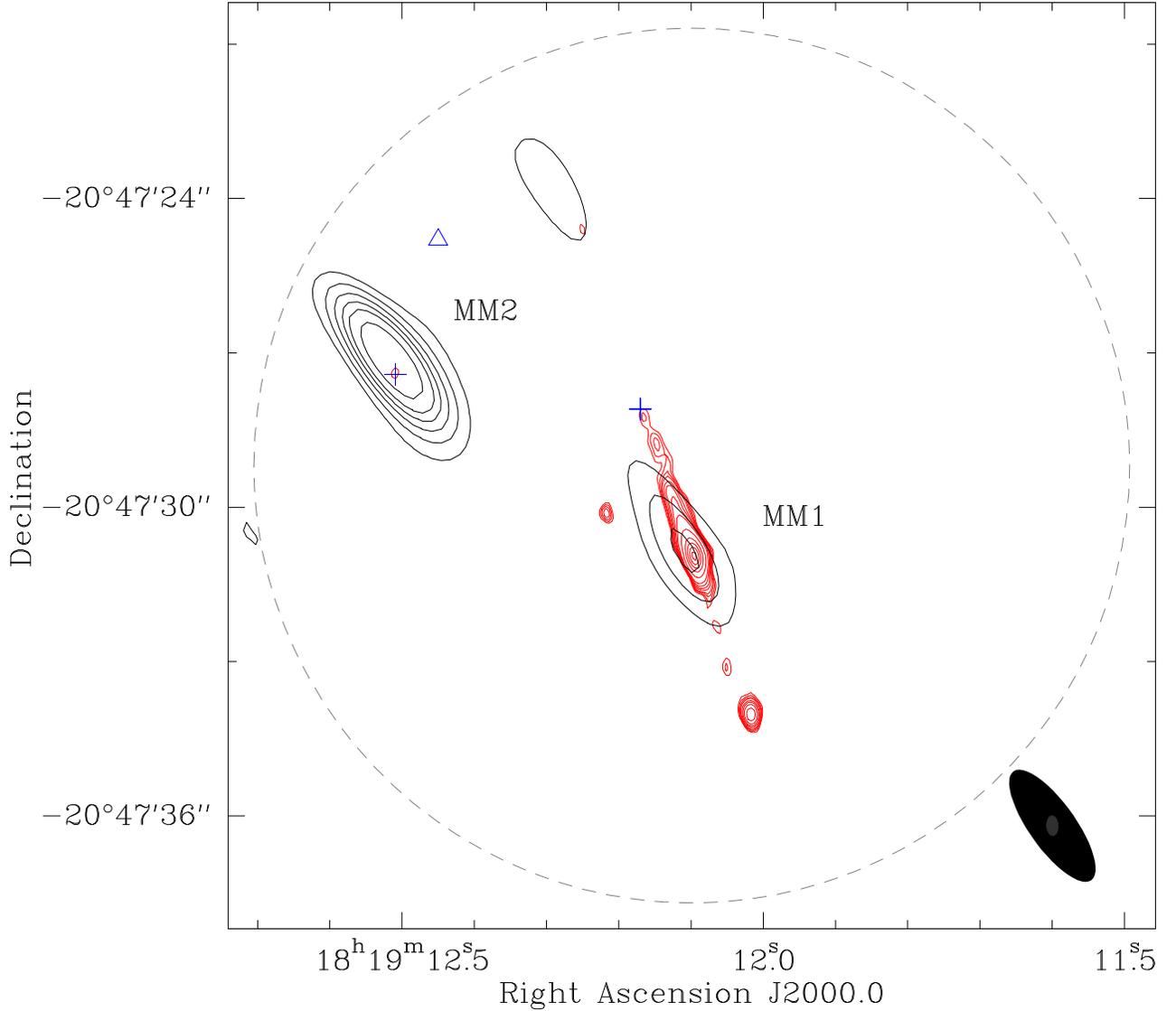}
\caption{Overlay of our 456~$\mu$m (658~GHz) continuum image (black contours)
with the 3.5~cm continuum VLA image (red contours) obtained by
\markcite{1995ApJ...453..268G}{G{\'o}mez} {et~al.} (1995) towards the central
region of GGD27. Contour levels are -6, -3, 3, 6, 9, 12, 15, 20, 30, 40 $\times
457$~mJy beam$^{-1}$, the rms noise of the 456~$\mu$m image and -5, 5, 6, 7, 8,
10, 12, 15, 20, 40,80, 120, 160, 200 $\times 9.1$~$\mu$Jy~beam$^{-1}$, the rms
noise of the 3.5~cm image (the latter, as appeared in the original article). The
456~$\mu$m image is presented without primary beam correction. The synthesized
beams are shown at the bottom right corner and the primary beam is indicated by
a dashed circle with radius $8\farcs5$ centered on the phase center. Symbols are
the same as in Fig. \ref{fig: 220g}. The dust emission towards MM1 coincides
with the radio continuum jet, the suspected powering source of the HH80-81
bipolar system. Likewise, the dust emission in MM2 coincides with a very weak
radio continuum source (the smallest contour around the cross shows the
5~$\sigma$ emission at 3.5~cm). The weak source to the north, near the primary
beam edge is an artifact of the cleaning process. 
}
\label{fig: 650g}
\end{figure}

Figure \ref{fig: 220g} presents the 1.36~mm (220~GHz) continuum emission images towards GGD27 obtained with the SMA in its compact and very extended configurations. The low angular resolution image shows a strong and elongated source, with a main peak and a possible second peak towards the NE. The high angular resolution image clearly separates the continuum emission into two sources, a strong and compact one, MM1, located near the phase center and a weak and extended one, MM2, located about 7$\arcsec$ to the NE from MM1. The 456~$\mu$m (658~GHz) image (see Fig. \ref{fig: 650g}) also shows two compact sources spatially coincident with MM1 and MM2 (see Table \ref{tab: T2}). MM1 dominates the emission at 1.36~mm and coincides spatially with the powerful radio continuum jet found in this region (\markcite{1993ApJ...416..208M,1995ApJ...449..184M,1995ApJ...453..268G}{Mart\'{\i}} {et~al.} 1993, 1995; {G{\'o}mez} {et~al.} 1995). MM2 spatially coincides (within $0\farcs1$) with a weak radio continuum source and a water maser (\markcite{1995ApJ...453..268G,1999A&A...345L...5M,2005AJ....130..711K}{G{\'o}mez} {et~al.} 1995; {Mart{\'{\i}}}, {Rodr{\'{\i}}guez}, \&  {Torrelles} 1999; {Kurtz} \& {Hofner} 2005).

Table \ref{tab: T2} shows the results of the Gaussian fits (position, peak flux, flux density and deconvolved size) carried out with the IMFIT task of AIPS in the three images, for MM1 and MM2. The image at 1.36~mm obtained with the compact configuration (left panel of Fig. \ref{fig: 220g}), shows a single extended source, which was fitted simultaneously with two Gaussians. This fit gives fluxes of 610$\pm$8~mJy and 443$\pm$9~mJy for MM1 and MM2, respectively and deconvolved sizes of $3\farcs5\pm0\farcs1\times1\farcs3\pm0\farcs2$; P.A.=$176\degr\pm3\degr$ for MM1 and $4\farcs9\pm0\farcs2\times1\farcs5\pm0\farcs5$; P.A.=$164\degr\pm3\degr$ for MM2. In the very extended configuration image (right panel of Fig. \ref{fig: 220g}), MM1 has a flux density of $441\pm6$~mJy and appears unresolved with an upper limit to its diameter of $0\farcs36$ (see Table \ref{tab: T2} and section \S~4 for the derivation of a more stringent upper limit). MM2 splits into two main components separated by $\sim 1\arcsec$. The eastern component, MM2(E), is compact and has a flux density of 57$\pm$6~mJy, with an upper limit to its diameter of $0\farcs36$. The western component, MM2(W), is slightly weaker (50$\pm$10~mJy) and extended, with a deconvolved size of $1\farcs1\pm0\farcs2\times0\farcs5\pm0\farcs1$; P.A.=$34\degr\pm11\degr$. In the 456~$\mu$m image (Fig. \ref{fig: 650g}) the detected sources are unresolved (diameter smaller than $1\farcs7$), with fluxes of 5.3$\pm$1.7~Jy for MM1 and 19$\pm$1.5~Jy for MM2.

\begin{figure}[h]
\centering
\includegraphics[width=0.8\textwidth]{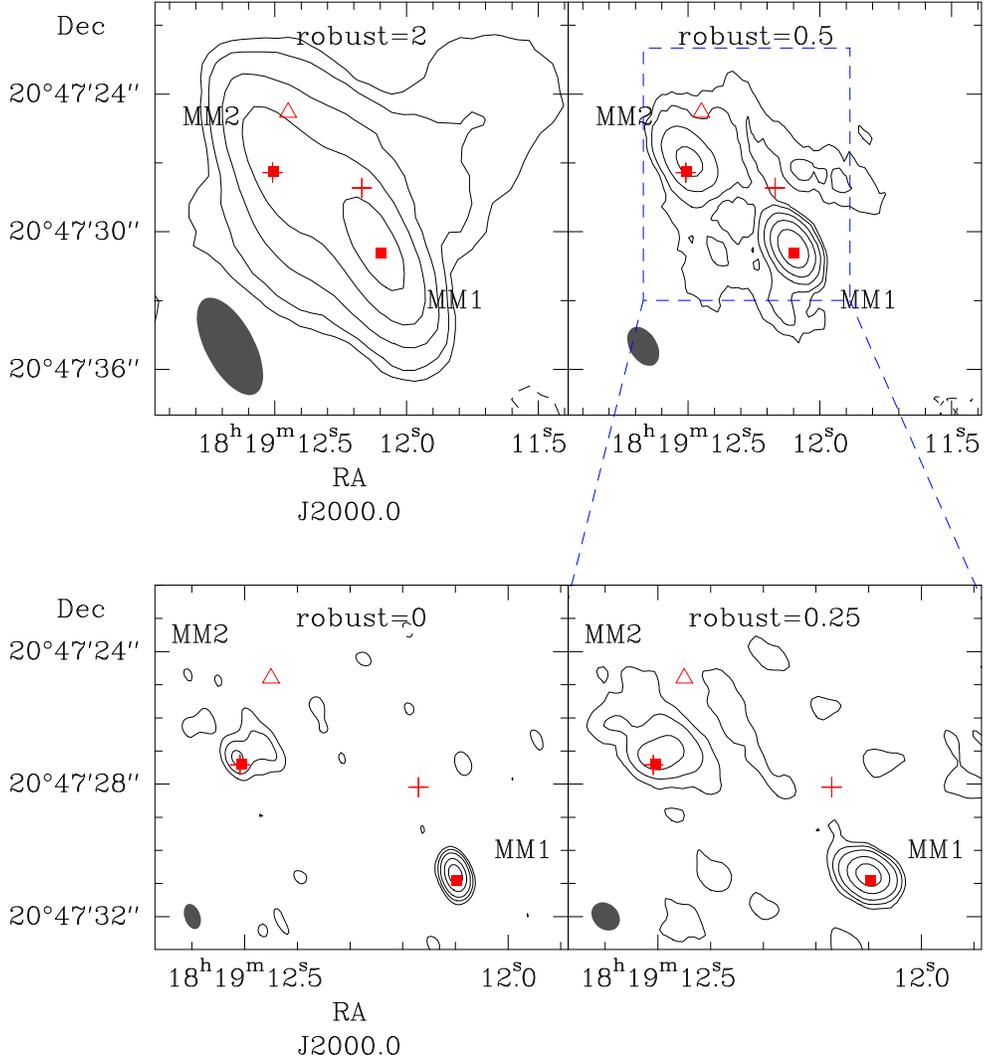}
\caption{Combined images of the extended and compact configuration at 1.36~mm (220~GHz) obtained with different values of the robust parameter. The value of the robust weighting is indicated within each panel. Contours are -5, -3, 3, 5, 10, 20 and 40 $\times 8.1$, $\times 5.3$~mJy beam$^{-1}$, the rms noise of the images with robust 2 and 0.5 (top panels), and -6, -3, 3, 6, 12, 30 and 60 $\times 4.2$ and $\times 3.9$~mJy beam$^{-1}$, the rms noise of the images with robust 0.25 and 0 (bottom panels), respectively. Symbols are the same as in Fig. \ref{fig: 220g}. Synthesized beams are $5\farcs0\times2\farcs4$ ($36\fdg3$), $2\farcs0\times1\farcs3$ ($45\fdg0$), $0\farcs9\times0\farcs7$ ($31\fdg0$) and $0\farcs8\times0\farcs4$ ($15\fdg1$) for the images with robust 2, 0.5, 0.25 and 0, respectively, and are shown as gray solid ellipses at the bottom left corner of each image.
}
\label{fig: F3}
\end{figure}

The total flux measured in the low angular resolution image at 1.36~mm (1.20~Jy) is slightly greater (by $\sim0.15$~Jy) than the sum of the fluxes of MM1 and MM2 obtained by fitting two Gaussians. This could suggest the existence of a weak extended component surrounding both sources. In addition, the total flux is greater than that measured in the high angular resolution image (550~mJy), which indicates that the extended emission in the low angular resolution image is resolved out in the high angular resolution image. In order to analyze the nature of this extended emission, the data at 1.36~mm of both configurations were combined (Fig. \ref{fig: F3}). The low and high angular resolution data overlaps in a uv-plane region between 18 and 52 k$\lambda$, in which their fluxes concide within 3\%. This allowed a straightforward combination of the data, using the INVERT routine of Miriad; the resultant images were obtained using several robust parameters (2, 0.5, 0.25 and 0) to sample intermediate angular resolutions between the images of compact and very extended configurations.
In the robust=2 image (see Fig. \ref{fig: F3}), the emissions from MM1 and MM2 are blended due to the synthesized beam size. However, when angular resolution improves, the emission from each source is clearly distinguished. Comparing the different angular resolution images, the flux and size of MM1 and MM2 decrease when angular resolution increases. In the images with robust$\leq$0.5 (synthesized beam size of about 1$\arcsec$), MM1 is unresolved and MM2 appears elongated in an E-W direction joining MM2(E) and MM2(W). A two-Gaussian fit on MM2 in the images with robust$\leq$0.25 shows MM2(E) marginally resolved ($\lesssim0.8\arcsec$) with a P.A. near to $50\degr\pm15\degr$, while in the robust=0 image it appears unresolved. On the other hand, MM2(W) appears spatially resolved in both images (robust=0.25 and robust=0 images), with sizes and orientations similar to those observed in the very extended configuration image (see Table \ref{tab: T2}). The two components of MM2 are only well resolved in the highest angular resolution image (image with robust=0).

The fluxes of MM1 and MM2 decrease in a different way with the angular resolution of the images. We note here that the area of the synthesized beam changes approximately a factor of 4 between the images with robust=2 and robust=0.5, and a factor of 8 between the images with robust=0.5 and robust=0 (see the caption of Fig. \ref{fig: F3}). The flux of MM1 decreases by 22\% between the robust=2 and the robust=0.5 images, and by only 6\% between the robust=0.5 and the robust=0 images. Thus, the MM1 flux decreases by 28\% (170~mJy) in all, with respect to the flux obtained with the robust=2 image. In the case of MM2, the flux decreases by 37\% between the images with robust 2 and 0.5, by 27\% between those with robust 0.5 and 0.25, and by 12\% between those with robust 0.25 and 0. Considering the flux of MM2(E) and MM2(W) altogether (each one contributes about half of the total flux measured in MM2), the total decrease of the MM2 flux is of about 76\% (340~mJy). Finally, these images also show about 135~mJy of extended emission, which is seen as weak emission bridges between MM1 and MM2 (see for instance the image with robust=0.5, Fig. \ref{fig: F3}), and as a weak filamentary structure towards the west of MM2 (Fig. \ref{fig: F3}, image with robust=0.25).

The flux variations with resolution observed in MM1 and MM2 indicate the existence of extended emission associated with these sources. These results also indicate that the emission of MM1 comes mainly from a compact component embedded in a weak extended envelope. In the case of MM2, most of the emission is extended and elongated in an East-West direction; the extended envelope surrounds the compact MM2(E), and the resolved source MM2(W). 
In section \S5 we discuss the morphology and the flux variations of the sources with the angular resolution as a possible evidence of MM1 and MM2 being in different evolutionary stages.

GGD27 has been previously observed at 1.4~mm with the BIMA and at 1.33~mm with the SMA interferometers (\markcite{2003ApJ...597..414G}{G{\'o}mez} {et~al.} 2003; \markcite{2009ApJ...702L..66Q}{Qiu} \& {Zhang} 2009) and at 450~$\mu$m with the JCMT (\markcite{1995MNRAS.276.1024J,1995AJ....110.1762M,2006A&A...453.1003T}{Jenness}, {Scott}, \&  {Padman} 1995; {McCutcheon} {et~al.} 1995; {Thompson} {et~al.} 2006). From BIMA maps (using a similar angular resolution as our compact configuration image), \markcite{2003ApJ...597..414G}{G{\'o}mez} {et~al.} (2003) reported a total flux of 1.18~Jy, in good agreement with the flux measured here in the compact configuration image. \markcite{2009ApJ...702L..66Q}{Qiu} \& {Zhang} (2009) reported a total flux of 2~Jy measured with the SMA at 1.33~mm, which is a factor of 2 greater than that reported here. The data used by \markcite{2009ApJ...702L..66Q}{Qiu} \& {Zhang} (2009) included SMA observations with compact and subcompact configurations, and obtained a better uv-coverage and a greater sensitivity to extended structures. The different JCMT observations at 450~$\mu$m showed a dusty core of $\sim40\arcsec$ of diameter nearly centered on MM1 and MM2. The flux obtained by the JCMT observations is of 97$\pm$30~Jy (\markcite{2006A&A...453.1003T}{Thompson} {et~al.} 2006), while the total flux recovered in our SMA 456~$\mu$m image is 24.3~Jy, which is 25\% of the flux measured with the JCMT. Since the SMA in its compact configuration is sensitive to structures smaller than 22$\arcsec$ at 1.36~mm and 7$\arcsec$ at 456~$\mu$m, the detected sources at both wavelengths correspond to the emission of compact structures located inside the core detected with the JCMT.

The SMA observations that we present here show that at 1.36~mm the continuum flux of MM1 is higher than that obtained for MM2, while at 456~$\mu$m, MM2 is the strongest source in the region. The spectral indices of MM1 and MM2, using a power-law model between 1.36~mm and 456~$\mu$m, are 2.3$\pm$0.5 and 3.6$\pm$0.4, respectively. We find that the free-free contribution based on 3.6~cm emission extrapolated to 1.36~mm is negligible ($\lesssim$5~mJy for MM1 and $\lesssim$0.5~mJy for MM2 at 1.36~mm, assuming free-free spectral indices of 0.2 and 0.6 respectively; see \S 3.2 below). Up to now, MM1 has been considered as the main source in the region and thus the one that contributes most of the bolometric luminosity, obtained from IRAS and low angular resolution (mostly single-dish) observations (see Fig. \ref{fig: F5}). However, the 456~~$\mu$m SMA observations suggest that MM2 contributes substantially to the total luminosity of the region.

\subsection{Spectral Energy Distributions}

The high angular resolution data obtained with the SMA towards GGD27, presented in this paper, allow the extension of the Spectral Energy Distributon (SED) of the sources MM1 and MM2. MM1 has been observed with high angular resolution in a very wide range of wavelengths (from radio to NIR; see Table \ref{tab: T4}). 
MM2, on the other hand, has only been detected at three different wavelengths: the two reported here and at 3.5~cm by \markcite{1995ApJ...453..268G}{G{\'o}mez} {et~al.} (1995).
In this section we analyze the nature of both sources by fitting the SED of MM1 and deriving the spectral index of MM2 in the (sub)millimeter range.

\begin{figure}[h]
\epsscale{0.7}
\plotone{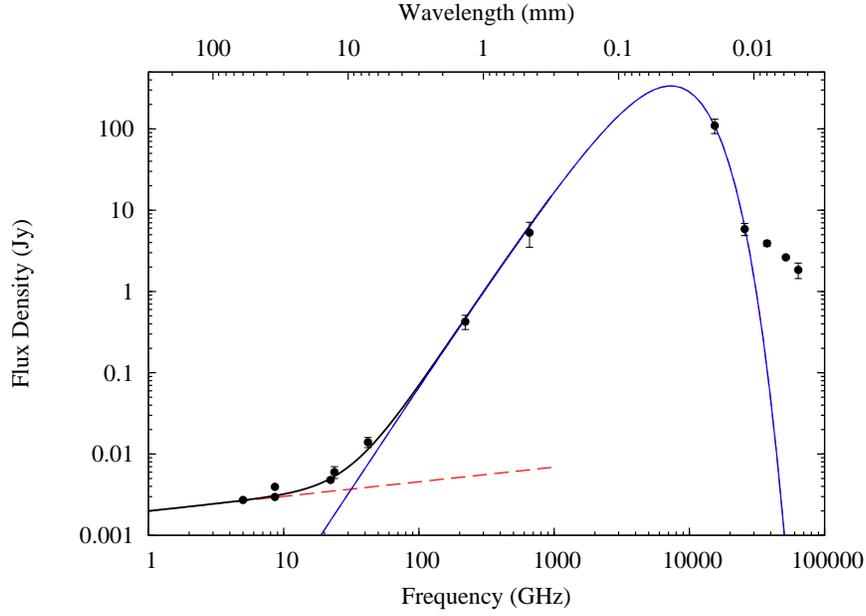}
\caption{\footnotesize{Spectral Energy Distribution from radio to NIR
wavelengths for MM1 (filled circles with error bars). The point-dashed curve
is the free-free emission and the blue curve the modified black body fit of
MM1's SED (see text). The three NIR points taken from Table \ref{tab: T4}, with
wavelengths below 10~$\mu$m, are not included in the fit due to the uncertainty
on the $\beta$ parameter of an additional modified black body.}
}
\label{fig: F4}
\end{figure}

\begin{figure}[h]
\epsscale{0.7}
\plotone{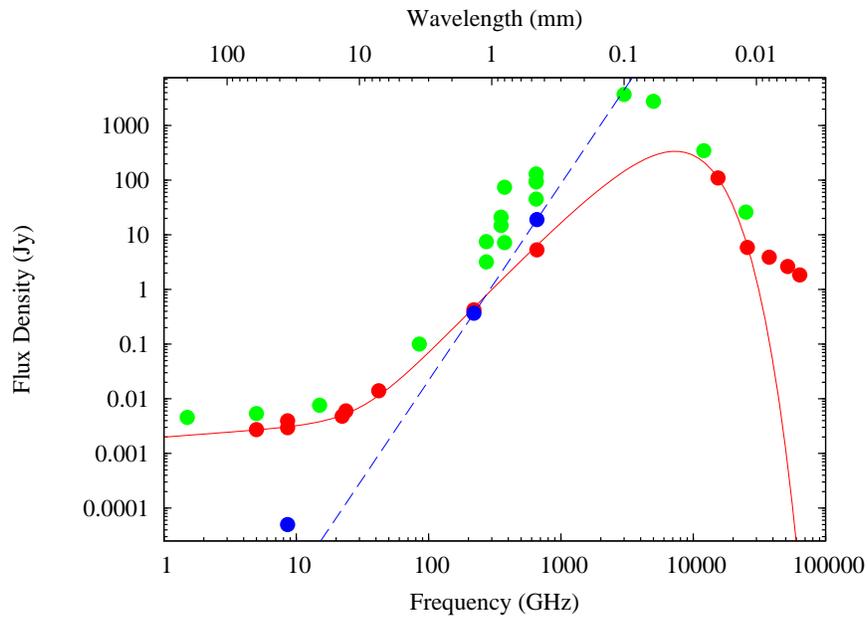}
\caption{\footnotesize{Spectral Energy Distribution from radio to NIR
wavelengths for MM1 (red circles) and for MM2 (blue circles). The red solid
curve accounts for the fit of MM1's SED (see text and Fig. \ref{fig: F4}). The
blue dashed curve is a power law adjusted to the mm and submm points of MM2 with
spectral index 3.6, corresponding to a $\beta$ of 1.6 (see section \S3.2). The
green circles correspond to the low resolution data taken from literature
(single-dish and IRAS measurements). Most of the error bars are smaller than the
size of each symbol.}
}
\label{fig: F5}
\end{figure}

The SED of MM1 was built using the data from radio (VLA) and (sub)millimeter (BIMA and SMA) interferometric observations, and from infrared observations with good angular resolution (see Table \ref{tab: T4}). The selected data have an angular resolution comparable to our 456~$\mu$m observations (i.e., about 2$\arcsec$). In this way, the measured emission comes from the vicinity of MM1, avoiding the contamination due to emission from other sources. Figure \ref{fig: F4} shows the SED of MM1 and a fit to these points. The fit minimizes the $\chi^2$ following the algorithms described in \markcite{2009A&A...501.1259C}{Cant{\'o}}, {Curiel}, \&  {Mart{\'{\i}}nez-G{\'o}mez} (2009). The function fitted consists of a modified black body plus free-free emission model:
$$S_{\nu} = B_{\nu}(T_{dust}) (1-e^{-\tau_{\nu}})\Omega_{s} + a_{\nu} \nu^{\alpha}\enskip,$$
where $B_{\nu}$ is the Plank function with a temperature T$_{dust}$, $\Omega_{s}$ is the solid angle subtended by MM1, $\tau_{\nu}$ is the opacity which depends on frequency as $\tau_{\nu} = \tau_0\cdot \left(\nu/\nu_0\right)^{\beta}$, $\beta$ being the dust opacity index and $\nu_0$ an arbitrary reference frequency, and $a_{\nu}=a_0/\nu^{\alpha}_0$. 
Since this model includes a single modified black body, the points with $\lambda<$~10~$\mu$m shown in Figure \ref{fig: F4} were not taken into account in the fit. We could fit these NIR data with an additional modified black body, but the uncertainty on the $\beta$ parameter prevents us from using a second component. However, these NIR data points seem to be associated with dust emission at high temperatures. Table \ref{tab: T5} shows the results of the best fit with 6-independent parameters ($T_{dust}$, $\tau_0$, $\beta$, $\Omega_{s}$, $a_0$ and $\alpha$), as well as the uncertainties provided by the algorithm. The obtained spectral index of the free-free emission ($\alpha=0.18\pm0.01$) is consistent with the value of $\alpha=0.2$ obtained by \markcite{1993ApJ...416..208M}{Mart\'{\i}} {et~al.} (1993) and by \markcite{2003ApJ...597..414G}{G{\'o}mez} {et~al.} (2003) independently, implying a negligible free-free contribution in the (sub)millimeter range. The obtained dust opacity index ($\beta=0.53\pm0.05$), agrees well with the value reported by \markcite{2003ApJ...597..414G}{G{\'o}mez} {et~al.} (2003) ($\beta=0.6$), obtained from a power-law fit from the flux densities obtained at 7, 3.5 and 1.4~mm. The low value of the dust opacity index suggests the presence of processed dust grains in MM1 (\markcite{2006ApJ...636.1114D}{Draine} 2006 and reference therein). In addition, the dust temperature obtained for MM1 is $109\pm3$~K and the estimated size of the source is $1.1\pm0.4$ arcsec$^2$, consistent with the compact nature of this source. Finally, the bolometric luminosity of MM1 was estimated as 3300~L$_{\sun}$. This luminosity is smaller than the luminosity obtained from low angular resolution data ($\sim2\times10^4$~L$_{\sun}$; see Fig. \ref{fig: F5}), and it could be considered as a lower limit, since the high angular resolution observations used in this fit do not include the outer parts of the main core (only includes the dense structure surrounding MM1), which are also heated by this source (see also the discussion in section \S 4).

In the case of MM2, there are only three high angular resolution ($<1\arcsec$) flux estimates (Fig. \ref{fig: F5}): the fluxes at 1.36~mm and 456~$\mu$m (this paper), and the flux at 3.5~cm (\markcite{1995ApJ...453..268G}{G{\'o}mez} {et~al.} 1995). The spectral index between 1.36~mm and 456~$\mu$m ($\alpha = 3.60\pm0.4$) corresponds to a dust opacity index $\beta$ between 1.2 and 2.0 (accounting for the error bars) which is similar to the averaged opacity index found in the interstellar medium ($\beta=1.6$; Draine 2006). This could indicate that the dust in MM2 contains mainly small particles, such as those found in the interstellar medium, which suggests that MM2 could be a very young source. In addition, the 3.5~cm emission is consistent with free-free emission, since the expected dust emission at this wavelength (using a spectral index of 3.6) is only 6\% of the measured radio continuum flux. This free-free emission could be arising from an ionized outflow associated with a young intermediate/high-mass protostar.

\subsection{Masses and Column Densities}
The equations used to estimate the mass and the H$_2$ column density from the continuum dust emission, expressed in appropriate units are (\markcite{1983QJRAS..24..267H}{Hildebrand} 1983): 
\begin{equation}
M[M_{\sun}] = 3.2643\cdot 10^6\cdot R \cdot \frac{(e^{0.048 \frac{\nu}{T_d}}-1)\cdot D^2\cdot S_{\nu}}{\nu^3\cdot k_{1.3mm}\cdot\nu^{\beta}}
\end{equation}
\begin{equation}
N_{H_2} [\mathrm{cm}^{-2}] = 8.4208\cdot 10^{23}\cdot \frac{M [\mathrm{M}_{\sun}]}{D^2\cdot\Omega}
\end{equation}
where $R$, $T_d$, $D$, $S_{\nu}$, $k_{\nu}$ and $\Omega$ are the gas-to-dust ratio (which is assumed to be 100), the dust temperature in Kelvin, the distance to the source in kpc (adopted as 1.7~kpc), the measured flux density in Jy, the dust opacity in $\mathrm{cm^2\cdot g}^{-1}$ and the angular size in arcsec$^2$. The frequency $\nu$ is in GHz. In order to obtain the dust opacity, $k_{1.3mm}=1.0$~cm$^2\cdot$g$^{-1}$ was assumed for thin ice mantles (\markcite{1994A&A...291..943O}{Ossenkopf} \& {Henning} 1994), scaling this value with frequency as $\nu^{\beta}$, where $\beta$ is the dust opacity index derived in \S 3.2 for MM1 and MM2. The hypothesis of thin ice mantles towards GGD27 is suggested by the detection of absorption features of water ices at 3 and 6~$\mu$m, and CO$_2$ ices at 4.3 and 15.2~$\mu$m (\markcite{2002A&A...381..571P,2004ApJ...611..928V}{Peeters} {et~al.} 2002; {van Diedenhoven} {et~al.} 2004). 
The mass, column density and number density were obtained for each source (see Table \ref{tab: T6}), from the measured flux densities at 1.36~mm (high and low angular resolution images) and at 456~$\mu$m. For MM1, we use T$_{dust}$=109~K and $\beta$=0.53 derived from the SED fit in \S 3.2, while for MM2 we adopt $\beta$=1.6 (see \S 3.2) and T$_{dust}$=35~K as derived from observations of several CH$_3$CN~(12-11) transitions associated with this source (\markcite{2009ApJ...702L..66Q}{Qiu} \& {Zhang} 2009). 
The uncertainties were estimated from the errors in T$_{dust}$, $\beta$ and the flux densities. These uncertainties do not take into account the effects introduced by the opacity, which could increase the value of the derived masses.

Using the flux densities obtained with the two-Gaussian fit in the low angular resolution image at 1.36~mm, the masses estimated for MM1 and MM2 are 5.6$\pm$0.2~M$_{\sun}$ and 14.9$\pm$0.4~M$_{\sun}$, respectively.
As expected, these values are similar to those obtained using the 456~$\mu$m flux densities (3$\pm$1~M$_{\sun}$ for MM1 and 17$\pm$2~M$_{\sun}$ for MM2).
On the other hand, from the high angular resolution data at 1.36~mm, we estimate that the mass of MM1 is 4.1$\pm$0.2~M$_{\sun}$ and the masses of MM2(E) and MM2(W) are 1.9$\pm$0.2~M$_{\sun}$ and 1.7$\pm$0.3~M$_{\sun}$, respectively. 
Thus, the extended dust emission surrounding MM1 and MM2 contains about 14~M$_{\sun}$, most of which ($\sim12$~M$_{\sun}$) seems to be associated with MM2. The protostars in MM2 still have a huge reservoir of gas and dust to accrete, while the gas and dust supply for the protostar in MM1 is almost an order of magnitude smaller (see discussion in section \S 4).

The total mass obtained from the fluxes in the low angular resolution image ($\sim$20~M$_{\sun}$) is, approximately, 5\% of the total mass obtained for the $\sim$arcmin dusty core detected by single-dish observations  (e.g., \markcite{2000AJ....119.2711H}{Hunter} {et~al.} 2000), within which MM1 and MM2 are embedded. Thus, only a small fraction of the material in the large scale core is associated with the compact structures. This suggests that the main core still has enough material to form a cluster of intermediate/high-mass protostars or to contribute with more accretion into MM1 and MM2. 

Finally, based on the high angular resolution observations at 1.36~mm, we obtain a lower limit for the column density of MM1 ($1.13\times 10^{25}$~cm$^{-2}$), MM2(E) ($5.5\times 10^{24}$~cm$^{-2}$) and MM2(W) ($8.9\times 10^{23}$~cm$^{-2}$). In addition, we also estimate lower limits for the volume number densities: $1.85\times 10^{9}$~cm$^{-3}$ for MM1, $9.0\times 10^{8}$~cm$^{-3}$ for MM2(E) and $1.1\times 10^{8}$~cm$^{-3}$ for MM2(W). Hence, the volume number density towards MM1 and MM2(E) could be one order of magnitude greater than in MM2(W). Such high densities ($\gtrsim10^{8}$~cm$^{-3}$) have also been found in other high-mass star formation regions (e.g. IRAS20126, \markcite{2005A&A...434.1039C}{Cesaroni} {et~al.} 2005; Cep A, \markcite{2007ApJ...661L.187J}{Jim{\'e}nez-Serra} {et~al.} 2007; W51 North, \markcite{2009ApJ...698.1422Z}{Zapata} {et~al.} 2009), around circumstellar structures (possible accretion disks).

\section{DISCUSSION}

The high angular resolution observations at 1.36~mm towards GGD27 show that MM1 is a compact source, which spatially coincides with the powerful radio continuum jet observed in this region (\markcite{1993ApJ...416..208M,1995ApJ...449..184M,1995ApJ...453..268G}{Mart\'{\i}} {et~al.} 1993, 1995; {G{\'o}mez} {et~al.} 1995). MM1 appears unresolved after applying a uniform weighting to the very extended configuration data at 1.36~mm. The SNR of MM1 in this image is $\sim100$ (see image with robust 0 in Fig. \ref{fig: F3}), so taking into account some additional considerations, it is possible to obtain a more restrictive upper limit for the diameter of this source than the $0\farcs36$ (i.e., about 600~AU at 1.7~kpc) given in \S 3.1. To make this new estimation, the dust emission from MM1 is assumed to arise from a flat envelope in the shape of an accretion disk. This hypothesis is based on the presence of a radio jet and a molecular outflow asociated with MM1. Such a disk should be oriented perpendicular to the jet's axis, whose P.A. is 21$\degr$ (\markcite{1993ApJ...416..208M}{Mart\'{\i}} {et~al.} 1993).
On the other hand, the uniform-weighted image obtained with the extended configuration data has a synthesized beam of $0\farcs72\times 0\farcs31$, with the major axis of the synthesized beam (P.A.=$19\fdg4$) oriented almost in the same direction as the radio jet. Thus, the major axis of the synthesized beam follows the direction of the radio jet, while the minor axis of the synthesized beam follows the orientation of the putative accretion disk. Therefore, given the high SNR in MM1 and assuming that the compact dust emission comes from an accretion disk, a reasonable upper limit for the diameter of this disk would be half the minor axis of the synthesized beam, which is $0\farcs16$ or 272~AU at the assumed distance. If the disk hypothesis is right, this is one of the smallest disks found around a high-mass protostar. Some of the disks detected around protostars with $10^3-10^5$~L$_{\sun}$ have a radius ranging between 70-1000~AU and a mass between 1-20~M$_{\sun}$ (\markcite{2007prpl.conf..197C}{Cesaroni} {et~al.} (2007); \markcite{2005Natur.437..109P}{Patel} {et~al.} (2005); \markcite{2001Sci...292.1513S}{Shepherd}, {Claussen}, \&  {Kurtz} (2001); \markcite{2007_Rodriguez}{Rodr{\'{\i}}guez}, {Zapata}, \&  {Ho} (2007); \markcite{2009_Franco}{Franco-Hern{\'a}ndez} {et~al.} (2009); \markcite{2010_Galvan}{Galv{\'a}n-Madrid} {et~al.} (2010).).

In addition, although the diameter of the disk in MM1 is similar (perhaps somewhat greater) to that of disks around low-mass protostars, its mass ($\sim$4~M$_{\sun}$) is about two orders of magnitude larger than those found in low-mass protostars ($\sim0.03$~M$_{\sun}$; e.g, \markcite{2006ApJ...636L.137P}{Palau} {et~al.} 2006). These massive disks have been reported in other sources. The existence of a disk around MM1 would imply a star formation mechanism based on the accretion of circumstellar material.

The morphology of the northern source (MM2) varies with the angular resolution of the 1.36~mm images (Fig. \ref{fig: F3}). These images show that MM2 contains a compact core, MM2(E), and an extended structure, MM2(W), both surrounded by an extended and massive envelope. MM2(E) is probably a compact core with a central protostar as indicated by the presence of a weak radio continuum source and a water maser (\markcite{1995ApJ...453..268G,1999A&A...345L...5M,2005AJ....130..711K}{G{\'o}mez} {et~al.} 1995; {Mart{\'{\i}}} {et~al.} 1999; {Kurtz} \& {Hofner} 2005). MM2(W) is probably a prestellar core with enough mass to form an intermediate-high mass protostar (probably an early-B type). Spectral line observations with high angular resolution may show whether MM2(W) is collapsing or not. In addition, given the small projected distance between MM2(E) and MM2(W) (about 1600~AU), the two sources might be gravitationally bound. There is enough mass to allow orbital motions of $\lesssim1.5$~km~s$^{-1}$. Finally, MM2 is characterized by strong submillimeter emission (19~Jy at 456~$\mu$m), low temperature ($T\sim35$~K, \markcite{2009ApJ...702L..66Q}{Qiu} \& {Zhang} 2009), high density ($\gtrsim10^7$~cm$^{-3}$), a dust opacity index close to that of the interstellar medium ($\beta\sim1.6$), non-detection in the MIR and NIR, association with a young molecular outflow (\markcite{2009ApJ...702L..66Q}{Qiu} \& {Zhang} 2009) and a massive dusty envelope (11~M$_{\sun}$). These properties are similar to the characteristics of the Class 0 low-mass protostars (\markcite{1993ApJ...406..122A}{Andr\`{e}}, {Ward-Thompson}, \&  {Barsony} 1993; \markcite{1994ApJ...420..837A}{Andr\`{e}} \& {Montmerle} 1994). However, MM2 seems to be embedded in an envelope with a mass four times greater than the envelopes of Class 0 low-mass protostars (which have 1-4~M$_{\sun}$; e.g., NGC 1333 IRAS4A, \markcite{2006Sci...313..812G}{Girart}, {Rao}, \&  {Marrone} 2006; L723, \markcite{2009ApJ...694...56G}{Girart}, {Rao}, \&  {Estalella} 2009; IRAS16293A, \markcite{2009ApJ...707..921R}{Rao} {et~al.} 2009). The higher mass of the envelope could also be explained by multiple protostars within it.

MM1 and MM2 show different characteristics. MM2 has a $\beta\sim1.6$ similar to the interstellar medium (\markcite{2006ApJ...636.1114D}{Draine} 2006 and references therein), while MM1 has a $\beta\approx0.5$ which suggests more processed dust (\markcite{2000prpl.conf..533B,2006ApJ...636.1114D,2007prpl.conf..767N}{Beckwith}, {Henning}, \&  {Nakagawa} 2000; {Draine} 2006; {Natta} {et~al.} 2007). The temperature of MM1 is greater than that of MM2 and the free-free emission of MM1 is two orders of magnitude greater than that of MM2. In addition, the outflow associated with MM1 has a mass of 407-460~M$_{\sun}$ and a dynamic time scale of $\sim10^5$~yr (\markcite{1989ApJ...347..894Y}{Yamashita} {et~al.} 1989; \markcite{2001A&A...378..495R}{Ridge} \& {Moore} 2001), while the outflow associated with MM2 has a mass of 0.22 M$_{\sun}$ and a dynamic time scale of $\sim2\times10^3$~yr (\markcite{2009ApJ...702L..66Q}{Qiu} \& {Zhang} 2009). Finally, the mass of the envelope of MM1 is almost an order of magnitude smaller than that of MM2, which could indicate that the protostars in MM2 are at an earlier stage of accretion from the envelope as compared to MM1. All these characteristics are consistent with MM2 being in a younger evolutionary stage than MM1.

MM1 is the strongest source at 1.36~mm, while MM2 dominates the emission at 456~$\mu$m. Therefore the present SMA data indicate that both sources significantly contribute to the total luminosity of the region, unlike previous publications (e.g., \markcite{1989RMxAA..17...59R}{Rodr{\'{\i}}guez} \& {Reipurth} 1989; \markcite{1999A&A...345L...5M,2001ApJ...547..292M,2008ApJ...685.1005Q}{Mart{\'{\i}}} {et~al.} 1999; {Molinari}, {Noriega-Crespo}, \&  {Spinoglio} 2001; {Qiu} {et~al.} 2008), which pointed at MM1 as the source responsible for all the bolometric luminosity in the region. With the available data, it is unclear which is the more luminous source, but their SEDs (Fig. \ref{fig: F5}) suggest comparable contribution to the total bolometric luminosity (estimated as $\sim2\times10^4$~L$_{\sun}$ from IRAS, JCMT and CSO data; see Fig. \ref{fig: F5}). According to Table \ref{tab: T5} of \markcite{1998A&A...336..339M}{Molinari} {et~al.} (1998), a luminosity of $10^4$~L$_{\sun}$ would suggest a 13-18~M$_{\sun}$ protostar. 

The luminosity of MM1 estimated from the fit to its SED is 3300~L$_{\sun}$, which is smaller than half the bolometric luminosity of the region. However, it has been shown (e.g., \markcite{1999ApJ...525..330Y}{Yorke} \& {Bodenheimer} 1999) that the inclination of a disk-protostar system is an important parameter for the estimations of the luminosity from the SED. In this case, the radio jet associated with MM1 is almost in the plane of the sky (\markcite{1991A&A...252..299A,1995ApJ...449..184M}{Aspin} {et~al.} 1991; {Mart\'{\i}} {et~al.} 1995), which implies that the circumstellar disk-like structure is approximately edge-on. The large column density derived in the disk-like structure yields a visual extinction of at least $1.4\times10^4$. This implies that along the line of sight most of the mid-IR radiation is effectively absorbed by the disk. In terms of total luminosity, this suggests that a significant fraction of the radiation escapes through the outflow cavity (\markcite{1999ApJ...525..330Y}{Yorke} \& {Bodenheimer} 1999; \markcite{2009Sci...323..754K}{Krumholz} {et~al.} 2009). Therefore, the bolometric luminosity of the young star in MM1 is likely larger than the measured 3300~L$_{\sun}$, and it would likely correspond to a B1 ZAMS star. However, the presence of the massive disk and of the powerful outflow indicates that it is still in a very active accretion phase, so the protostar could become a B0 ZAMS star, or even a more massive star.

Finally, assuming that MM1 is excited by a B1-type protostar, it should exhibit a compact HII region. Free-free emission at several wavelengths has been reported towards this source, but the spectral index (0.18$\pm$0.01, section \S3.2) clearly corresponds to that expected for an ionized wind. This suggests that the protostar in MM1 has not yet developed an HII region and therefore it could be considered as a \textit{true} high-mass protostar.

\section{CONCLUSIONS}
The present paper presents a study of the GGD27 high-mass star formation region based on SMA observations of the continuum emission at 1.36~mm (compact and very extended configurations) and 456~$\mu$m (compact configuration). These observations show compact dust emission in two regions, MM1 and MM2. Using these new observations together with previously published results, we analyze the SEDs of MM1 and MM2. The masses and densities of these two sources are also estimated. The main conclusions of this paper are:

1. The dust continuum emission at 1.36~mm and 456~$\mu$m comes from two compact structures (MM1 and MM2), separated 7$\arcsec$ (about 12000~AU in projected distance) and embedded in a $\sim40\arcsec$ core, that was detected by single-dish observations. The high angular resolution observations at 1.36~mm show that MM1 is unresolved (with an estimated radius of less than 150~AU), with an estimated mass of 4.1~M$_{\sun}$. MM1 spatially coincides with an IRAS source of $2\times10^4$~L$_{\sun}$ and it is associated with a powerful molecular outflow and a radio continuum jet. These results suggest that the dust emission in MM1 probably arises from a massive accretion disk. At subarcsecond angular scales, MM2 splits into two main components, MM2(E) and MM2(W), that are separated by 1$\arcsec$ (1700 AU in projected distance), each with an estimated mass of $\sim2$~M$_{\sun}$. MM2(E) is probably a compact core with a young protostar inside (indicated by a weak radio continuum source and a water maser) and MM2(W) is probably a prestellar core. Both sources are embedded in a massive envelope of at least 11~M$_{\sun}$.

2. The characteristics of MM1 and MM2 are very different: (a) MM2 has a steeper spectral index at (sub)mm wavelengths, suggesting a less processed dust than in MM1, (b) MM1 has a higher temperature, suggesting that there is a higher mass protostar inside MM1 than in MM2, (c) the molecular outflow associated with MM2 is younger and less massive than the one associated with MM1, (d) MM1 is unresolved even at subarcsecond scales, while MM2 appears extended with two main components and faint emission surrounding them. These characteristics suggest that MM1 and MM2 are in different evolutionary stages, with MM2 younger than MM1.

3. The SMA observations show that both MM1 and MM2 contribute significantly to the bolometric luminosity of the whole region ($\sim2\times10^4$~L$_{\sun}$). A fit to the SED of MM1 indicates that a lower limit to the luminosity of this source is 3300~L$_{\sun}$, which implies that MM1 is at least a B1 ZAMS protostar. However, the likely continued accretion in MM1 indicates that this source will accrete several solar masses. At present it is not clear what kind of protostar is inside MM2. However, these SMA observations suggest that MM2 contributes an important fraction of the bolometric luminosity of this region.

We thank all members of the SMA staff that made these observations possible. The authors are very grateful to Todd Hunter and Jun-Hui Zhao for assistance in the calibration of the data. 
MFL acknowledges financial support from DGAP-UNAM, M\'exico. SC and MFL acknowledge support from CONACyT grant 60581, M\'exico. SC and MFL thank the hospitality of the Institut de Ciencies de l'Espai (CSIC-IEEC), Bellaterra, Catalunya, Spain. JMG is supported by Spanish MICINN grant AYA2008-06189-C03 and Catalan AGAUR grant 2009 SGR1172. YG acknowledges support from CONACyT grants 80769 and 49947-F.

\appendix

\section{APPENDIX}
The phase transfer calibration method relies on the fact that phase variations at high frequencies may be correlated in some manner to those at low frequencies, where the SMA is more stable and more sensitive (the phase noise from the system and the sky are much better when working at such low frequencies). Since the atmosphere is almost non-dispersive (i.e., its refraction index does not depend strongly on frequency, away from resonant lines), to first order, the ratio of phases at two different frequencies is expected to be linear with the frequency ratio (i.e., $\propto\nu_1/\nu_2$, see e.g. \markcite{2005stt..conf...58H}{Hunter} {et~al.} 2005). Phase transfer is expected to be useful with submillimeter interferometers as the SMA, due to the lack of strong calibrators at high frequencies, which could make the standard techniques of phase calibration unavailable. 

\begin{figure}[h]
\epsscale{1}
\plotone{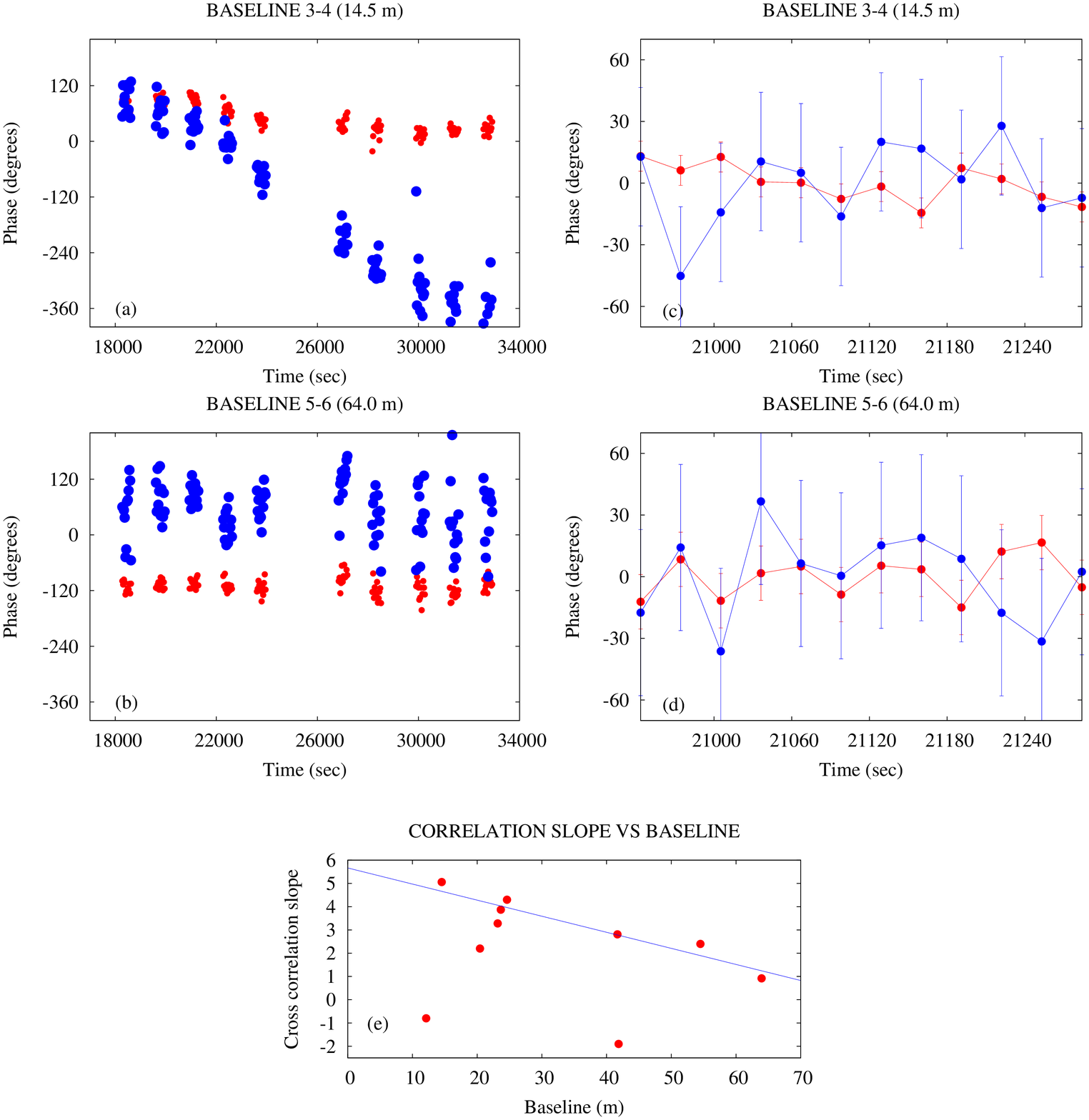}
\caption{\footnotesize{(a) and (b): 215~GHz (red circles) and 658~GHz
(blue circles) uncalibrated phase of the masers of VX Sgr from two baselines
with very different separations (the 3-4 baseline with 14.5 m and the 5-6
baseline with 64.0 m). Each point represents the average value of a 30 seconds
integration. (c) and (d): 5 minutes of the phase behaviour at 215~GHz (red
circles) and 658~GHz (blue circles) for both baselines, after subtracting a
first order polynomial fitted to the raw data. Error bars are derived from the
rms noise of this 5 minutes data. (e): cross correlation slope versus the
baseline separation. The straight line is a fit to most of the measurements
(filled circles) but the correlations of baselines 2-3, 3-5 and 4-6 (which are
located well below the general trend). }
}
\label{fig: F6}
\end{figure}

VX Sgr, the evolved star included in the observing cycle, has maser emission from SiO at 215.596~GHz and from H$_2$O at 658.007~GHz. Both masers were detected simultaneously by the SMA, so we analysed their phase behavior in order to check the phase transfer procedure.
In first place, the beginning and the end of the track were flagged due to poor quality (when elevation of GGD27 and VX Sgr were below $30\degr$). The data were corrected for variations in system temperature and bandpass, and as in \S 2.2.2 we made a pseudo-continuum channel for each maser. The phases of the masers showed in general good linear correlations (above an 80\%). These correlations gave different values for the slope and the abscisa offset depending on the baseline lengths (panels (a), (b) and (e) of Fig. \ref{fig: F6}). The linear dependence of the slopes of the correlation with the length of each baseline suggests that there are instrumental effects between the antennas and the correlator. On the other hand, the phase drifts seem to be almost linear during long periods of time. Hence, after self calibrate each set of maser data with a time averaging of 10 minutes, we removed these slow drifts. 
After that we tried to cross-correlate the phases of both masers again, but now they were uncorrelated. The panels (c) and (d) of Figure \ref{fig: F6} show the behaviour of the phases of both masers for the two baselines. Ideally, we expect the phases to be linearly related by the ratio between frequencies, but in this case, the rms of the phase at 658~GHz (about 40$\degr$), is probably masking the theoretical behaviour. This would occur at high frequencies because of the higher rms of the phases.  

Despite the high rms of the phase at high frequency we tested the phase transfer technique. We scaled up the 215~GHz phase gain solutions to that of the 658~GHz uncalibrated data of VX Sgr, using the Miriad task PHATRANS. PHATRANS extrapolated the phase corrections from the phase gains determined at 215~GHz with the SiO maser, to the phase corrections at 658~GHz of the VX Sgr uncalibrated data, using the linear relation between the phase gains of the SiO and water pseudo-continuum channels at both frequencies. Then, we performed the flux calibration using Ceres. After continuum subtraction, the output cleaned and restored velocity cube image of the water maser at 658~GHz (spectral resolution of 0.37~km~s$^{-1}$) had an rms of 18~Jy~beam$^{-1}$, and a SNR of 22. Five iterations of self calibration on this velocity cube, improved its SNR. The final velocity cube has an rms of 16~Jy~beam$^{-1}$ and a SNR of 36.

To check the validity of the phase transfer calibration, we also performed a similar calibration of the data of VX Sgr at 658~GHz applying directly the phase gain solutions of the  the pseudo-continuum channel of the water maser. We refer to this calibration technique as phase referencing. The only difference with the phase transfer procedure was the phase calibration step. In this case, the final velocity cube image of the maser had an rms of 16~Jy~beam$^{-1}$ and a SNR of 38. Self calibration did not improve this image substantially (less than 0.5 in SNR). The phase transfer calibration of the water maser of VX Sgr shows a similar output data than the phase referencing calibration: a) The SNR of the velocity cubes obtained with phase transfer and phase referencing are similar (36 and 38, respectively). b) Both velocity cubes have the same qualitative information (i.e., the same channels show emission over $3~\sigma$). c) The spectra obtained with the two calibration techniques have qualitatively the similar shape (same width and peak position). The difference between both spectra is smaller than $3~\sigma$ for practically all the spectral channels. d) The moment 0 images are similar qualitatively, showing an unresolved source at the phase center. e) The total flux measured in the moment 0 images is only $5\%$ smaller in the phase transfer image. The SNR of the phase tranfer and the phase referencing images are 19 and 20, respectively. Thus, the analysis indicates that in this case, phase transfer could give a similar output data (images and spectrum) than the phase referencing, taking into account the rms of the images and the calibration uncertainties.

We also tried to apply the phase transfer procedure to the GGD 27 data. However, in this case, the source was undetected, preventing to start with self calibration. Therefore, our attempt to apply the phase transfer technique seems to be quite successful when calibrating the bright water maser of VX Sgr, but it failed with a fainter source such as the continuum emission of GGD 27. We can conclude that the phase transfer technique described here works, at least in this case, for a source strong enough to apply further self calibration. 

\newpage
%% \bibliography

\newpage

%%%%%%%%%%%%%%%%%%%%%%%%%%%%%%%%%%%%%%%%%%%%%%%%%%%%%%%%%%%%%%%%%%%%%%
%FIGURES
%%%%%%%%%%%%%%%%%%%%%%%%%%%%%%%%%%%%%%%%%%%%%%%%%%%%%%%%%%%%%%%%%%%%%%

%%%%%%%%%%%%%%%%%%%%%%%%%%%%%%%%%%%%%%%%%%%%%%%%%%%%%%%%%%%%%%%%%%%%%%
%TABLES
%%%%%%%%%%%%%%%%%%%%%%%%%%%%%%%%%%%%%%%%%%%%%%%%%%%%%%%%%%%%%%%%%%%%%%

\begin{deluxetable}{lcccccc}
\tablewidth{0pt}
\tablecaption{Continuum observations}
\tablehead{
\colhead{Continuum} &\colhead{Frequency}  &\colhead{Date} & \multicolumn{2}{c}{Synthesized beam}  & \colhead{Primary beam}   &  \colhead{rms} \\
\colhead{Map} & \colhead{}  & \colhead{}  & \colhead{HPBW} & \colhead{P.A.} & \colhead{}  & \colhead{} \\
\colhead{} & \colhead{(GHz)}  & \colhead{}  & \colhead{($\arcsec$)} & \colhead{($\degr$)} & \colhead{($\arcsec$)}  & \colhead{(mJy beam$^{-1}$)} \\
}
\startdata
Lower Side Band & 215 & 2005 Aug 24 & $8.3\times 3.1$ & $34.2$ & 50 & 11 \\
UpperSide Band  & 225 & 2005 Aug 24 & $7.9\times 3.0$ & $34.1$ & 50 & 10 \\
LSB + USB      & 220 & 2005 Aug 24 & $8.1\times 3.0$ & $34.1$ & 50 & 10 \\
Lower Side Band & 215 & 2007 May 29 & $0.70\times 0.39$ & $13.2$ & 50 & 4.3 \\
UpperSide Band  & 225 & 2007 May 29 & $0.67\times 0.36$ & $12.8$ & 50 & 4.1 \\
LSB + USB      & 220 & 2007 May 29 & $0.68\times 0.38$ & $13.0$ & 50 & 3.0 \\
UpperSide Band\tablenotemark{(a)} & 658 & 2005 Aug 24 & $2.4\times 1.1$ & $37.1$ & 17 & 678 \\
\enddata 
\tablenotetext{(a)}{Primary beam corrected image.}
\label{tab: T1}
\end{deluxetable}

\begin{deluxetable}{lcccccc}
\tablewidth{0pt}
\tablecaption{Flux calibration. Planets, moons and other calibrators.}
\tabletypesize{\scriptsize}
\tablehead{
\colhead{Source} &\colhead{Frequency} & \colhead{Measured Flux\tablenotemark{(a)}} & \colhead{Expected Flux\tablenotemark{(b)}} & \colhead{Deviation} & \colhead{Offset\tablenotemark{(c)}} & \colhead{Observing Date} \\
\colhead{} & \colhead{(GHz)} & \colhead{(Jy)} & \colhead{(Jy)} & \colhead{(\%)} & \colhead{($\arcsec$)} & \colhead{}
}
\startdata
J1911-201 & 	215 & 	2.13 & 	\nodata\tablenotemark{(d)} & \nodata & 0.024,-0.011 & Aug 24,2005 \\
J1911-201 & 	225 & 	2.18 & 	2.37 & 	8.0 & 	0.020,-0.008 & 	Aug 24,2005 \\
3C454.3 & 	215 & 	12.87 & \nodata\tablenotemark{(d)} & \nodata & 0.010,0.045 & Aug 24,2005 \\
3C454.3 & 	225 & 	28.27 & 	27.85 & 1.5 & 0.003,0.030 & 	Aug 24,2005 \\
Ganymede & 	215 & 	3.06 & 	3.52 & 	12.9 & 0.022,0.015 & Aug 24,2005 \\
Ganymede & 	225 & 	3.29 & 	3.84 & 	14.3 & 0.010,0.013 & 	Aug 24,2005 \\
Callisto\tablenotemark{(e)} & 	215 & 	3.34 & 	3.57 & 	6.2 & -0.003,-0.010 & 	Aug 24,2005 \\
Callisto\tablenotemark{(e)} & 	225 & 	3.63 & 	3.90 & 	6.7 & 0.005,0.014 & 	Aug 24,2005 \\
Neptune & 	215 & 	13.80 & 13.41 & 3.0 & -0.019,-0.001 & 	Aug 24,2005 \\
Neptune & 	225 & 	14.21 & 13.93 & 2.4 & 0.014,0.003 & Aug 24,2005 \\
Ceres & 	215 & 	0.83 & 	0.99 & 	15.9 & 0.030,0.016 & 	Aug 24,2005 \\
Ceres & 	225 & 	0.83 & 	1.08 & 	23.0 & 0.005,0.012 & Aug 24,2005 \\
\tableline
J1911-201 & 	658 & 	\nodata\tablenotemark{(f)} & $<1.5$ & \nodata & \nodata & Aug 24,2005 \\
3C454.3 & 	658 & 	13.50 & \nodata\tablenotemark{(d)} & \nodata & -0.004,0.013 & Aug 24,2005 \\
Ganymede & 	658 & 	25.82 & 29.40 & 12.4 & 	-0.057,-0.057 & Aug 24,2005 \\
Callisto & 	658 & 	32.38 & 30.38 & 6.4 & 	0.028,0.054 & Aug 24,2005 \\
Neptune & 	658 & 	46.30 & 68.84 & 32.8 & 	-0.008,0.056 & Aug 24,2005 \\
Ceres\tablenotemark{(e)} & 	658 & 	9.03 & 	8.67 & 3.9 & 	0.032,0.009 & Aug 24,2005 \\
\tableline
J1911-201\tablenotemark{(e)} & 	215 & 	1.37 & 	1.24 & 	10.5 & 	0.004,-0.004 & 	May 29, 2007 \\
J1911-201\tablenotemark{(e)} & 	225 & 	1.29 & 	1.16 & 	11.2 & 	0.004,-0.002 & 	May 29, 2007 \\
3C454.3\tablenotemark{(g)} & 	220 & 	7.69 & 	7.41 & 	3.8 & 	0.000,0.002 & 	May 29, 2007 \\
mwc349 & 	215 & 	1.44 & 	\nodata\tablenotemark{(d)} & \nodata & 0.002,-0.003 & May 29, 2007 \\
mwc349 & 	225 & 	1.92 & 	1.73 & 	11.0 & 	-0.000,-0.000 & May 29, 2007 \\
\enddata 
\tablenotetext{(a)}{Flux density within a box surrounding the emission from the calibrator.}
\tablenotetext{(b)}{Data interpolated from the SMA monitoring program of calibrators and from the \textit{planetary visibility function calculator} of the SMA webpage (Gurwell, M.).}
\tablenotetext{(c)}{Peak position from a fit performed with the MAXFIT task of Miriad. The offsets are relative to the phase center of each source.}
\tablenotetext{(d)}{Data not available from the SMA monitoring program of calibrators.}
\tablenotetext{(e)}{Flux calibrator.}
\tablenotetext{(f)}{Undetected at this frequency.}
\tablenotetext{(g)}{The observed data at 220~GHz is an average from data at 215 and 225~GHz (7.69 and 9.27 Jy respectively). This value is compared with that measured by the SMA monitoring program at 219.8~GHz on the same day of our observations.}
\label{tab: T9}
\end{deluxetable}

\begin{deluxetable}{cccccr@{$\pm$}lr@{$\pm$}lcr@{$\pm$}l}
\tablewidth{0pt}
\tablecolumns{8}
\tabletypesize{\scriptsize}
\tablecaption{Source parameters}
\phs
\tablehead{
\colhead{Frequency} & \colhead{Beam} & \colhead{Source} & \colhead{RA\tablenotemark{(a)}} & \colhead{DEC\tablenotemark{(a)}} & \multicolumn{2}{c}{S$_{peak}$} & \multicolumn{2}{c}{S$_{int}$} &  \colhead{Size\tablenotemark{(b)}} &  \multicolumn{2}{c}{P.A.} \\
\colhead{(GHz)} & \colhead{} & \colhead{}  & \colhead{$18^h19^m$} & \colhead{$-20\degr47\arcmin$} & \multicolumn{2}{c}{(mJy beam$^{-1}$)}  & \multicolumn{2}{c}{(mJy)} & \colhead{($\arcsec$)} & \multicolumn{2}{c}{($\degr$)}
}
\startdata
220   & $8\farcs1\times 3\farcs0$; $34\fdg1$ & MM1\tablenotemark{(c)} & $12\fs114$ & $30\farcs61$ & 455&4  & 610&8 & $3.5\pm0.1\times 1.3\pm0.2$    & 176&3 \\
220   & $8\farcs1\times 3\farcs0$; $34\fdg1$ & MM2\tablenotemark{(c)} & $12\fs488$ & $27\farcs30$  & 260&4  & 443&9     & $4.9\pm0.2\times 1.5\pm0.5$   & 164&3   \\
658   & $2\farcs4\times 1\farcs1$; $37\fdg1$ & MM1 & $12\fs105$ & $30\farcs90$ & 4900&900 & 5300&1700 & $\lesssim1.17$ & \multicolumn{2}{c}{\nodata}  \\
658   & $2\farcs4\times 1\farcs1$; $37\fdg1$ & MM2 & $12\fs517$ & $27\farcs09$ & 18600&900 & 19000&1500 & $\lesssim1.17$ & \multicolumn{2}{c}{\nodata} \\
220   & $0\farcs68\times 0\farcs38$; $13\fdg0$ & MM1\tablenotemark{(d)} & $12\fs099$ & $30\farcs77$ & 409&3 & 441&6     & $\lesssim0.16$\tablenotemark{(e)}  & \multicolumn{2}{c}{\nodata}  \\
220  & $0\farcs68\times 0\farcs38$; $13\fdg0$ & MM2(E)\tablenotemark{(d)} & $12\fs515$ & $27\farcs23$ & 47&3  & 57&6    & $\lesssim0.36$ & \multicolumn{2}{c}{\nodata}  \\
220  & $0\farcs68\times 0\farcs38$; $13\fdg0$ & MM2(W)\tablenotemark{(d)} & $12\fs456$ & $27\farcs00$ & 17&3  & 50&10    & $1.1\pm0.2\times 0.5\pm0.1$ & 34&11  \\
\enddata 
\tablecomments{Natural weighting was applied in all the images. Fits were carried out with the IMFIT routine of AIPS. The errors are obtained from the fits and do not include the 20\% of the calibration uncertainties. The fluxes from the image at 456$\mu$m are primary beam corrected.}
\tablenotetext{(a)}{J2000.}
\tablenotetext{(b)}{Deconvolved sizes. When deconvolution fails we use half the synthesized beam size area ($(1/2)\cdot(\pi b_{maj}\cdot b_{min}/4)$) as an upper limit for the size of the region and calculate its diameter as $\sqrt{2\cdot\mbox{Beam area}/\pi}$. This could be a useful effective diameter for elliptical synthesized beams with high eccentricity.}
\tablenotetext{(c)}{In the low angular resolution image, the fit was carried out with two Gaussians simultaneously, because MM1 and MM2 share part of their emission. 
The uncertainties on the flux density and the sizes could be higher than the given by the fitting algorithm because the two sources are not well resolved.}
\tablenotetext{(d)}{In the high angular resolution image, the fit was carried out with one Gaussian for MM1 and independently, two Gaussians for MM2.}
\tablenotetext{(e)}{See \S 4 for the explanation of this upper limit.}
\label{tab: T2}
\end{deluxetable}

\begin{deluxetable}{ccccc}
\tablewidth{0pt}
\tablecaption{MM1 SED data}
\tablehead{
\colhead{Wavelength} & \colhead{Integrated Flux} & \colhead{Error} & \colhead{Aperture} & \colhead{Reference} \\
\colhead{mm} & \colhead{Jy} & \colhead{Jy} & \colhead{$\arcsec$} & \colhead{}
}
\startdata
60 & 0.00272 & 0.00002 & $0.8\times0.5$ & 1 \\
36 & 0.00296 & 0.00002 & $0.5\times0.3$ & 1 \\
35 & 0.00395 & 0.00005 & $0.4\times0.24$ & 2 \\
13 & 0.0048 & 0.00002 & $0.1\times0.1$ & 3 \\
13 & 0.006 & 0.001 & $4\times4$ & 4 \\
7 & 0.014 & 0.002 & $0.4\times0.4$ & 4 \\
1.36\tablenotemark{(a)} & 0.425 & 0.085 & $1.8\times1.3$ & 5 \\
0.46 & 5.3 & 1.8 & $2.4\times1.1$ & 5 \\
0.0195 & 109.6 & 21.9\tablenotemark{(b)} & $2.3\times2.3$ & 6 \\
0.0117 & 5.86 & 1.17\tablenotemark{(b)} & $2.3\times2.3$ & 6 \\
0.0080\tablenotemark{(b)} & 3.9 & 0.3\tablenotemark{(c)} & $2.4\times2.4$ & 7 \\
0.0058\tablenotemark{(b)} & 2.63 & 0.1\tablenotemark{(c)} & $2.4\times2.4$ & 7 \\
0.0047\tablenotemark{(b)} & 1.8 & 0.36\tablenotemark{(c)} & $2.3\times2.3$ & 6 \\
\enddata 
\tablerefs{(1) \markcite{1993ApJ...416..208M}{Mart\'{\i}} {et~al.} (1993); (2) \markcite{1995ApJ...453..268G}{G{\'o}mez} {et~al.} (1995); (3) \markcite{1999A&A...345L...5M}{Mart{\'{\i}}} {et~al.} (1999); (4) \markcite{2003ApJ...597..414G}{G{\'o}mez} {et~al.} (2003); (5) this work; (6) \markcite{1994A&A...292L...9A}{Aspin} {et~al.} (1994); (7) \markcite{2008ApJ...685.1005Q}{Qiu} {et~al.} (2008).}
\tablenotetext{(a)}{The data at this wavelength were obtained by combining our high and low angular resolution data  and applying the weighting factor, robust=0.5. This image yields a beam area similar to that for the image at 456~$\mu$m.}
\tablenotetext{(b)}{Points not included in the fit of the SED of MM1.}
\tablenotetext{(c)}{The errors were estimated to be 20\%.}
\label{tab: T4}
\end{deluxetable}

\begin{deluxetable}{cccccccc}
\tablewidth{0pt}
\tablecaption{MM1 SED fit}
\tablehead{
\colhead{Source} & \colhead{$\Omega_s$} & \colhead{$\tau_{456\mu m}$} & \colhead{$\beta$} & \colhead{$T_{dust}$} & \colhead{$a_{456\mu m}$\tablenotemark{(a)}} & \colhead{$\alpha$} & \colhead{Luminosity} \\
\colhead{} & \colhead{$(\arcsec)^ 2$} & \colhead{} & \colhead{} & \colhead{K} & \colhead{Jy} & \colhead{} & \colhead{$L_{\sun}$}
}
\startdata
MM1 & 1.1$\pm0.4$ & 0.22$\pm 0.08$ & 0.53$\pm 0.05$ & 109$\pm 3$ & $0.0064\pm 0.0004$ & 0.18$\pm0.01$ & 3300$\pm$500 \\
\enddata 
\tablenotetext{(a)}{Free-free emission contribution at 456~$\mu$m in Jy.}
\label{tab: T5}
\end{deluxetable}

\begin{deluxetable}{ccccccccc}
\tablewidth{0pt}
\tabletypesize{\scriptsize}
\tablecaption{Estimated masses and column densities.}
\tablehead{
\colhead{Source} & \colhead{Beam} & \colhead{Frequency} & \colhead{$k_{\nu}$} & \colhead{T$_{dust}$} & \colhead{Diameter} & \colhead{$M$} & \colhead{N$_{H_2}$\tablenotemark{(a)}} & \colhead{n\tablenotemark{(b)}} \\
\colhead{} & \colhead{} & \colhead{(GHz)}  & \colhead{(cm$^{2}$ g$^{-1}$)} & \colhead{K} & \colhead{$\arcsec$} & \colhead{(M$_{\sun}$)} & \colhead{($\times 10^{23}$ cm$^{-2}$)}  &  \colhead{($\times 10^{7}$ cm$^{-3}$)} 
}
\startdata
MM1 & $0\farcs68\times 0\farcs38$; $13\fdg0$ & 220   & 0.97 & 109 & 0.36 & 4.1$\pm$0.2 & 113 & 185 \\
MM1 & $8\farcs1\times 3\farcs0$; $34\fdg1$ & 220  & 0.97 & 109 & 2.1 & 5.6$\pm$0.2 & 4.7$\pm$0.3 & 1.33 \\
MM1 & $2\farcs4\times 1\farcs1$; $37\fdg1$ & 658  & 1.74 & 109 & 1.17 & 3$\pm$1 & 9 & 5  \\
\tableline
MM2(E) & $0\farcs68\times 0\farcs38$; $13\fdg0$ & 220  & 0.93 & 35 & 0.36 & 1.9$\pm$0.2 & 55.0 & 90 \\
MM2(W) & $0\farcs68\times 0\farcs38$; $13\fdg0$ & 220  & 0.93 & 35 & 0.7 & 1.7$\pm$0.3 & 13$\pm$1  & 11 \\
MM2    & $8\farcs1\times 3\farcs0$; $34\fdg1$ & 220  & 0.93 & 35 & 2.7 & 14.9$\pm$0.4  & 7.6$\pm$0.4 & 1.7  \\
MM2    & $2\farcs4\times 1\farcs1$; $37\fdg1$ & 658  & 5.35 & 35 & 1.17 & 17$\pm$2  & 46.7 & 24  \\
\enddata 
\tablenotetext{(a)}{Some column densities are lower limits because the unresolved nature of the sources and do not include uncertainties (the same applies for number densities).}
\tablenotetext{(b)}{It is assumed a spherical shape of the sources for the number density estimations.}
\label{tab: T6}
\end{deluxetable}

\begin{deluxetable}{lcccccccc}
\tablewidth{0pt}
\tablecaption{Properties of the millimeter sources}
\tablehead{
\colhead{Source} & \colhead{$\beta$\tablenotemark{(a)}} & \colhead{R} & \colhead{N$_{H_2}$} & \colhead{M$_{500}$\tablenotemark{(b)}} & \colhead{M$_{2500}$\tablenotemark{(c)}} & \colhead{T} & \colhead{L} & \colhead{t$_{dyn}$\tablenotemark{(d)}} \\
\colhead{} & \colhead{} & \colhead{(AU)} & \colhead{(cm$^{-2}$)} & \colhead{(M$_{\sun}$)} & \colhead{(M$_{\sun}$)} & \colhead{(K)} & \colhead{(L$_{\sun}$)} & \colhead{yr}
}
\startdata
MM1    & 0.5 & $<136$     & $1.1\times10^{25}$ & 4.1 & 5.6 & 110 & $\geq3.3\times$10$^3$ & $10^5$ \\
MM2(E) & 1.6 & $<300$ & $5.5\times10^{24}$ & 1.9 & 14.9  & 35\tablenotemark{(e)}  & \nodata & $2\times10^3$ \\
MM2(W) & \tablenotemark{(f)} & $600$ & $1.3\times10^{24}$ & 1.7 & \tablenotemark{(f)}  & \tablenotemark{(f)}  & \nodata & \nodata\\
\enddata 
\tablenotetext{(a)}{Dust opacity index.}
\tablenotetext{(b)}{Mass of the dust inside a radius of 500~AU (from the high angular resolution observations).}
\tablenotetext{(c)}{Mass of the dust inside a radius of 2500~AU (from the low angular resolution observations).}
\tablenotetext{(d)}{The dynamical time of the outflows associated with MM1 and MM2(E) (\markcite{2009ApJ...702L..66Q}{Qiu} \& {Zhang} 2009).}
\tablenotetext{(e)}{Temperature given in \markcite{2009ApJ...702L..66Q}{Qiu} \& {Zhang} 2009.}
\tablenotetext{(f)}{These quantities are estimated for MM2(E) and MM2(W) together.}
\label{tab: T7}
\end{deluxetable}

% \begin{deluxetable}{lccccc}
% \tablewidth{0pt}
% \tablecaption{Possible disks towards High-Mass YSOs}
% \tablehead{
% \colhead{} & \colhead{D} & \colhead{L} & \colhead{M$_{disk}$} & \colhead{R$_{disk}$} & \colhead{Reference} \\
% \colhead{} & \colhead{(pc)} & \colhead{(L$_{\sun}$)} & \colhead{(M$_{\sun}$)} & \colhead{(AU)} & \colhead{}
% }
% \startdata
% GGD27-MM1 & 1700 & $\geq3.3\times$10$^3$ & 4.1 & $<136$ & (1) \\
% Orion BN & 450 & 2.5$\times$10$^3$-10$^4$ & ? & 500 & (2) \\
% Orion I & 450 & 4$\times$10$^3$-10$^5$ & ? & 500 & (2) \\
% IRAS 20126+4104 & 1700 & 10$^4$ & 4 & 1600 & (2) \\
% Cep A HW2 & 725 & $\sim$10$^4$ & 1-8 & 330 & (3) \\
% AFGL 2591 & 1000 & 2$\times$10$^4$ & 0.4-1.8 & 500 & (2) \\
% G192.16-3.82 & 2000 &  2$\times$10$^3$ & 3-20  & 70 & (4) \\
% W51 & 6000 & 3$\times$10$^6$ & 40 & 1500 & (5) \\
% NGC 6334 I(N) & 1700 & 1.9$\times$10$^3$ & $\sim3$ & 1500 & (6) \\
% IRAS 16547-4247 & 2900 & 6$\times$10$^4$ & 5 & 1800 & (7) \\
% W33A & 3800 & 1$\times$10$^5$ & $\sim5$ & 1000 & (8)
% \enddata 
% \tablerefs{(1) from this work; (2) \markcite{2007prpl.conf..197C}{Cesaroni} {et~al.} (2007); (3) \markcite{2005Natur.437..109P}{Patel} {et~al.} (2005); (4) \markcite{2001Sci...292.1513S}{Shepherd}, {Claussen}, \&  {Kurtz} (2001); (5) \markcite{2009ApJ...698.1422Z}{Zapata} {et~al.} (2009); (6)\markcite{2007_Rodriguez}{Rodr{\'{\i}}guez}, {Zapata}, \&  {Ho} (2007); (7) \markcite{2009_Franco}{Franco-Hern{\'a}ndez} {et~al.} (2009); (8) \markcite{2010_Galvan}{Galv{\'a}n-Madrid} {et~al.} (2010).}
% \label{tab: T7}
% \end{deluxetable}


\begin{thebibliography}{}

\bibitem[{Andr\`{e}} \& {Montmerle} 1994]{1994ApJ...420..837A}
{Andr\`{e}}, P. \& {Montmerle}, T. 1994, \apj, 420, 837

\bibitem[{Andr\`{e}}, {Ward-Thompson}, \&  {Barsony} 1993]{1993ApJ...406..122A}
{Andr\`{e}}, P., {Ward-Thompson}, D., \& {Barsony}, M. 1993, \apj, 406, 122

\bibitem[{Aspin}, {Casali}, {Geballe}, \&  {McCaughrean} 1991]{1991A&A...252..299A}
{Aspin}, C., {Casali}, M.~M., {Geballe}, T.~R., \& {McCaughrean}, M.~J. 1991,  \aap, 252, 299

\bibitem[{Aspin} \& {Geballe} 1992]{1992A&A...266..219A}
{Aspin}, C. \& {Geballe}, T.~R. 1992, \aap, 266, 219

\bibitem[{Aspin}, {Puxley}, {Blanco}, {Pina}, {Pickup},  {Paterson}, {Sylvester}, {Laird}, {Bridger}, {Daly}, \&  {Griffin} 1994]{1994A&A...292L...9A}
{Aspin}, C., {Puxley}, P.~J., {Blanco}, P.~R., {Pina}, R.~K., {Pickup}, D.~A.,  {Paterson}, M.~J., {Sylvester}, J., {Laird}, D.~C., {et al.} 1994, \aap, 292, L9

\bibitem[{Beckwith}, {Henning}, \&  {Nakagawa} 2000]{2000prpl.conf..533B}
{Beckwith}, S.~V.~W., {Henning}, T., \& {Nakagawa}, Y. 2000, Protostars and  Planets IV, 533

\bibitem[{Beltr{\'a}n}, {Cesaroni}, {Codella},  {Testi}, {Furuya}, \& {Olmi} 2006]{2006Natur.443..427B}
{Beltr{\'a}n}, M.~T., {Cesaroni}, R., {Codella}, C., {Testi}, L., {Furuya},  R.~S., \& {Olmi}, L. 2006, \nat, 443, 427

\bibitem[{Benedettini}, {Molinari}, {Testi}, \&  {Noriega-Crespo} 2004]{2004MNRAS.347..295B}
{Benedettini}, M., {Molinari}, S., {Testi}, L., \& {Noriega-Crespo}, A. 2004,  \mnras, 347, 295

\bibitem[{Bonnell}, {Bate}, {Clarke}, \&  {Pringle} 2001]{2001MNRAS.323..785B}
{Bonnell}, I.~A., {Bate}, M.~R., {Clarke}, C.~J., \& {Pringle}, J.~E. 2001,  \mnras, 323, 785

\bibitem[{Bonnell}, {Bate}, \&  {Zinnecker} 1998]{1998MNRAS.298...93B}
{Bonnell}, I.~A., {Bate}, M.~R., \& {Zinnecker}, H. 1998, \mnras, 298, 93

\bibitem[{Briggs} 1995]{1995_Briggs}
{Briggs}, D.~S. 1995, in Bulletin of the American Astronomical Society,  Vol.~27, Bulletin of the American Astronomical Society, 1444

\bibitem[{Briggs}, {Schwab}, \& {Sramek} 1999]{1999_Briggs}
{Briggs}, D.~S., {Schwab}, F.~R., \& {Sramek}, R.~A. 1999, in Astronomical  Society of the Pacific Conference Series, Vol. 180, Synthesis Imaging in  Radio Astronomy II, ed. {G.~B.~Taylor, C.~L.~Carilli, \& R.~A.~Perley}, 127

\bibitem[{Cant{\'o}}, {Curiel}, \&  {Mart{\'{\i}}nez-G{\'o}mez} 2009]{2009A&A...501.1259C}
{Cant{\'o}}, J., {Curiel}, S., \& {Mart{\'{\i}}nez-G{\'o}mez}, E. 2009, \aap,  501, 1259

\bibitem[{Cesaroni}, {Galli}, {Lodato}, {Walmsley},  \& {Zhang} 2007]{2007prpl.conf..197C}
{Cesaroni}, R., {Galli}, D., {Lodato}, G., {Walmsley}, C.~M., \& {Zhang}, Q.  2007, Protostars and Planets V, 197

\bibitem[{Cesaroni}, {Neri}, {Olmi}, {Testi},  {Walmsley}, \& {Hofner} 2005]{2005A&A...434.1039C}
{Cesaroni}, R., {Neri}, R., {Olmi}, L., {Testi}, L., {Walmsley}, C.~M., \&  {Hofner}, P. 2005, \aap, 434, 1039

\bibitem[{Curiel} 1995]{1995RMxAC...1...59C}
{Curiel}, S. 1995, in Revista Mexicana de Astronomia y Astrofisica Conference  Series, Vol.~1, Revista Mexicana de Astronomia y Astrofisica Conference  Series, ed. {S.~Lizano \& J.~M.~Torrelles}, 59

\bibitem[{Draine} 2006]{2006ApJ...636.1114D}
{Draine}, B.~T. 2006, \apj, 636, 1114

\bibitem[{Franco-Hern{\'a}ndez}, {Moran},  {Rodr{\'{\i}}guez}, \& {Garay} 2009]{2009_Franco}
{Franco-Hern{\'a}ndez}, R., {Moran}, J.~M., {Rodr{\'{\i}}guez}, L.~F., \&  {Garay}, G. 2009, \apj, 701, 974

\bibitem[{Galv{\'a}n-Madrid}, {Zhang},  {Keto}, {Ho}, {Zapata}, {Rodr{\'{\i}}guez}, {Pineda}, \&  {V{\'a}zquez-Semadeni} 2010]{2010_Galvan}
{Galv{\'a}n-Madrid}, R., {Zhang}, Q., {Keto}, E., {Ho}, P.~T.~P., {Zapata},  L.~A., {Rodr{\'{\i}}guez}, L.~F., {Pineda}, J.~E., \& {V{\'a}zquez-Semadeni},  E. 2010, ArXiv e-prints

\bibitem[{Girart}, {Estalella}, {Viti}, {Williams}, \&  {Ho} 2001]{2001ApJ...562L..91G}
{Girart}, J.~M., {Estalella}, R., {Viti}, S., {Williams}, D.~A., \& {Ho},  P.~T.~P. 2001, \apjl, 562, L91

\bibitem[{Girart}, {Rao}, \&  {Estalella} 2009]{2009ApJ...694...56G}
{Girart}, J.~M., {Rao}, R., \& {Estalella}, R. 2009, \apj, 694, 56

\bibitem[{Girart}, {Rao}, \&  {Marrone} 2006]{2006Sci...313..812G}
{Girart}, J.~M., {Rao}, R., \& {Marrone}, D.~P. 2006, Science, 313, 812

\bibitem[{Girart}, {Rodr\'{\i}guez}, {Anglada},  {Estalella}, {Torrelles}, {Mart\'{\i}}, {Pena}, {Ayala}, {Curiel}, \&  {Noriega-Crespo} 1994]{1994ApJ...435L.145G}
{Girart}, J.~M., {Rodr\'{\i}guez}, L.~F., {Anglada}, G., {Estalella}, R.,  {Torrelles}, J.~M., {Mart\'{\i}}, J., {Pena}, M., {Ayala}, S., {et al.} 1994, \apjl, 435, L145

\bibitem[{G{\'o}mez}, {Rodr{\'{\i}}guez}, {Girart},  {Garay}, \& {Mart{\'{\i}}} 2003]{2003ApJ...597..414G}
{G{\'o}mez}, Y., {Rodr{\'{\i}}guez}, L.~F., {Girart}, J.~M., {Garay}, G., \&  {Mart{\'{\i}}}, J. 2003, \apj, 597, 414

\bibitem[{G{\'o}mez}, {Rodr\'{\i}guez}, \&  {Mart\'{\i}} 1995]{1995ApJ...453..268G}
{G{\'o}mez}, Y., {Rodr\'{\i}guez}, L.~F., \& {Mart\'{\i}}, J. 1995, \apj, 453,  268

\bibitem[{Heathcote}, {Reipurth}, \&  {Raga} 1998]{1998AJ....116.1940H}
{Heathcote}, S., {Reipurth}, B., \& {Raga}, A.~C. 1998, \aj, 116, 1940

\bibitem[{Hildebrand} 1983]{1983QJRAS..24..267H}
{Hildebrand}, R.~H. 1983, \qjras, 24, 267

\bibitem[{Ho}, {Moran}, \& {Lo} 2004]{2004ApJ...616L...1H}
{Ho}, P.~T.~P., {Moran}, J.~M., \& {Lo}, K.~Y. 2004, \apjl, 616, L1

\bibitem[{Hunter}, {Barrett}, {Blundell},  {Christensen}, {Kimberk}, {Leiker}, {Marrone}, {Paine}, {Cosmo Papa},  {Patel}, {Riddle}, {Smith}, {Sridharan}, {Tong}, {Young}, \&  {Zhao} 2005]{2005stt..conf...58H}
{Hunter}, T.~R., {Barrett}, J.~W., {Blundell}, R., {Christensen}, R.~D.,  {Kimberk}, R.~S., {Leiker}, S., {Marrone}, D.~P., {Paine}, S.~N., {et al.} 2005, in Sixteenth International  Symposium on Space Terahertz Technology, held May 2-4, 2005 at Chalmers  University of Technology. G{\"o}teborg, Sweden., p.58-63, 58--63

\bibitem[{Hunter}, {Churchwell}, {Watson}, {Cox},  {Benford}, \& {Roelfsema} 2000]{2000AJ....119.2711H}
{Hunter}, T.~R., {Churchwell}, E., {Watson}, C., {Cox}, P., {Benford}, D.~J.,  \& {Roelfsema}, P.~R. 2000, \aj, 119, 2711

\bibitem[{Jenness}, {Scott}, \&  {Padman} 1995]{1995MNRAS.276.1024J}
{Jenness}, T., {Scott}, P.~F., \& {Padman}, R. 1995, \mnras, 276, 1024

\bibitem[{Jijina} \& {Adams} 1996]{1996ApJ...462..874J}
{Jijina}, J. \& {Adams}, F.~C. 1996, \apj, 462, 874

\bibitem[{Jim{\'e}nez-Serra},  {Mart{\'{\i}}n-Pintado}, {Rodr{\'{\i}}guez-Franco}, {Chandler}, {Comito}, \&  {Schilke} 2007]{2007ApJ...661L.187J}
{Jim{\'e}nez-Serra}, I., {Mart{\'{\i}}n-Pintado}, J.,  {Rodr{\'{\i}}guez-Franco}, A., {Chandler}, C., {Comito}, C., \& {Schilke}, P.  2007, \apjl, 661, L187

\bibitem[{Kahn} 1974]{1974A&A....37..149K}
{Kahn}, F.~D. 1974, \aap, 37, 149

\bibitem[{Keto} \& {Wood} 2006]{2006ApJ...637..850K}
{Keto}, E. \& {Wood}, K. 2006, \apj, 637, 850

\bibitem[{Krumholz}, {Klein}, {McKee}, {Offner}, \&  {Cunningham} 2009]{2009Sci...323..754K}
{Krumholz}, M.~R., {Klein}, R.~I., {McKee}, C.~F., {Offner}, S.~S.~R., \&  {Cunningham}, A.~J. 2009, Science, 323, 754

\bibitem[{Kurtz} \& {Hofner} 2005]{2005AJ....130..711K}
{Kurtz}, S. \& {Hofner}, P. 2005, \aj, 130, 711

\bibitem[{Kurtz}, {Hofner}, \&  {{\'A}lvarez} 2004]{2004ApJS..155..149K}
{Kurtz}, S., {Hofner}, P., \& {{\'A}lvarez}, C.~V. 2004, \apjs, 155, 149

\bibitem[{Larson} \& {Starrfield} 1971]{1971A&A....13..190L}
{Larson}, R.~B. \& {Starrfield}, S. 1971, \aap, 13, 190

\bibitem[{Mart\'{\i}}, {Rodr\'{\i}guez}, \&  {Reipurth} 1993]{1993ApJ...416..208M}
{Mart\'{\i}}, J., {Rodr\'{\i}guez}, L.~F., \& {Reipurth}, B. 1993, \apj, 416,  208

\bibitem[{Mart\'{\i}}, {Rodr\'{\i}guez}, \&  {Reipurth} 1995]{1995ApJ...449..184M}
---. 1995, \apj, 449, 184

\bibitem[{Mart\'{\i}}, {Rodr\'{\i}guez}, \&  {Reipurth} 1998]{1998ApJ...502..337M}
---. 1998, \apj, 502, 337

\bibitem[{Mart{\'{\i}}}, {Rodr{\'{\i}}guez}, \&  {Torrelles} 1999]{1999A&A...345L...5M}
{Mart{\'{\i}}}, J., {Rodr{\'{\i}}guez}, L.~F., \& {Torrelles}, J.~M. 1999,  \aap, 345, L5

\bibitem[{Masqu{\'e}}, {Girart}, {Beltr{\'a}n},  {Estalella}, \& {Viti} 2009]{2009ApJ...695.1505M}
{Masqu{\'e}}, J.~M., {Girart}, J.~M., {Beltr{\'a}n}, M.~T., {Estalella}, R., \&  {Viti}, S. 2009, \apj, 695, 1505

\bibitem[{Masson} 1994]{1994ASPC...59...87M}
{Masson}, C.~R. 1994, in Astronomical Society of the Pacific Conference Series,  Vol.~59, IAU Colloq. 140: Astronomy with Millimeter and Submillimeter Wave  Interferometry, ed. {M.~Ishiguro \& J.~Welch}, 87

\bibitem[{McCutcheon}, {Sato}, {Purton},  {Matthews}, \& {Dewdney} 1995]{1995AJ....110.1762M}
{McCutcheon}, W.~H., {Sato}, T., {Purton}, C.~R., {Matthews}, H.~E., \&  {Dewdney}, P.~E. 1995, \aj, 110, 1762

\bibitem[{McKee} \& {Tan} 2003]{2003ApJ...585..850M}
{McKee}, C.~F. \& {Tan}, J.~C. 2003, \apj, 585, 850

\bibitem[{Molinari}, {Brand}, {Cesaroni}, {Palla}, \&  {Palumbo} 1998]{1998A&A...336..339M}
{Molinari}, S., {Brand}, J., {Cesaroni}, R., {Palla}, F., \& {Palumbo},  G.~G.~C. 1998, \aap, 336, 339

\bibitem[{Molinari}, {Noriega-Crespo}, \&  {Spinoglio} 2001]{2001ApJ...547..292M}
{Molinari}, S., {Noriega-Crespo}, A., \& {Spinoglio}, L. 2001, \apj, 547, 292

\bibitem[{Nakano} 1989]{1989ApJ...345..464N}
{Nakano}, T. 1989, \apj, 345, 464

\bibitem[{Natta}, {Testi}, {Calvet}, {Henning},  {Waters}, \& {Wilner} 2007]{2007prpl.conf..767N}
{Natta}, A., {Testi}, L., {Calvet}, N., {Henning}, T., {Waters}, R., \&  {Wilner}, D. 2007, Protostars and Planets V, 767

\bibitem[{Ossenkopf} \& {Henning} 1994]{1994A&A...291..943O}
{Ossenkopf}, V. \& {Henning}, T. 1994, \aap, 291, 943

\bibitem[{Palau}, {Ho}, {Zhang}, {Estalella}, {Hirano},  {Shang}, {Lee}, {Bourke}, {Beuther}, \& {Kuan} 2006]{2006ApJ...636L.137P}
{Palau}, A., {Ho}, P.~T.~P., {Zhang}, Q., {Estalella}, R., {Hirano}, N.,  {Shang}, H., {Lee}, C., {Bourke}, T.~L., {et al.} 2006,  \apjl, 636, L137

\bibitem[{Patel}, {Curiel}, {Sridharan}, {Zhang},  {Hunter}, {Ho}, {Torrelles}, {Moran}, {G{\'o}mez}, \&  {Anglada} 2005]{2005Natur.437..109P}
{Patel}, N.~A., {Curiel}, S., {Sridharan}, T.~K., {Zhang}, Q., {Hunter}, T.~R.,  {Ho}, P.~T.~P., {Torrelles}, J.~M., {Moran}, J.~M., {et al.} 2005, \nat, 437, 109

\bibitem[{Peeters}, {Mart{\'{\i}}n-Hern{\'a}ndez},  {Damour}, {Cox}, {Roelfsema}, {Baluteau}, {Tielens}, {Churchwell}, {Kessler},  {Mathis}, {Morisset}, \& {Schaerer} 2002]{2002A&A...381..571P}
{Peeters}, E., {Mart{\'{\i}}n-Hern{\'a}ndez}, N.~L., {Damour}, F., {Cox}, P.,  {Roelfsema}, P.~R., {Baluteau}, J., {Tielens}, A.~G.~G.~M., {Churchwell}, E., {et al.} 2002,  \aap, 381, 571

\bibitem[{Qiu} \& {Zhang} 2009]{2009ApJ...702L..66Q}
{Qiu}, K. \& {Zhang}, Q. 2009, \apjl, 702, L66

\bibitem[{Qiu}, {Zhang}, {Megeath}, {Gutermuth},  {Beuther}, {Shepherd}, {Sridharan}, {Testi}, \& {De  Pree} 2008]{2008ApJ...685.1005Q}
{Qiu}, K., {Zhang}, Q., {Megeath}, S.~T., {Gutermuth}, R.~A., {Beuther}, H.,  {Shepherd}, D.~S., {Sridharan}, T.~K., {Testi}, L., {et al.} 2008,  \apj, 685, 1005

\bibitem[{Rao}, {Girart}, {Marrone}, {Lai}, \&  {Schnee} 2009]{2009ApJ...707..921R}
{Rao}, R., {Girart}, J.~M., {Marrone}, D.~P., {Lai}, S., \& {Schnee}, S. 2009,  \apj, 707, 921

\bibitem[{Reipurth} \& {Graham} 1988]{1988A&A...202..219R}
{Reipurth}, B. \& {Graham}, J.~A. 1988, \aap, 202, 219

\bibitem[{Ridge} \& {Moore} 2001]{2001A&A...378..495R}
{Ridge}, N.~A. \& {Moore}, T.~J.~T. 2001, \aap, 378, 495

\bibitem[{Rodr\'{\i}guez}, {Moran},  {Dickinson}, \& {Giulbudagian} 1978]{1978ApJ...226..115R}
{Rodr\'{\i}guez}, L.~F., {Moran}, J.~M., {Dickinson}, D.~F., \& {Giulbudagian},  A.~L. 1978, \apj, 226, 115

\bibitem[{Rodr\'{\i}guez}, {Moran}, {Ho}, \&  {Gottlieb} 1980]{1980ApJ...235..845R}
{Rodr\'{\i}guez}, L.~F., {Moran}, J.~M., {Ho}, P.~T.~P., \& {Gottlieb}, E.~W.  1980, \apj, 235, 845

\bibitem[{Rodr{\'{\i}}guez} \& {Reipurth} 1989]{1989RMxAA..17...59R}
{Rodr{\'{\i}}guez}, L.~F. \& {Reipurth}, B. 1989, Revista Mexicana de  Astronomia y Astrofisica, 17, 59

\bibitem[{Rodr{\'{\i}}guez}, {Zapata}, \&  {Ho} 2007]{2007_Rodriguez}
{Rodr{\'{\i}}guez}, L.~F., {Zapata}, L.~A., \& {Ho}, P.~T.~P. 2007, \apjl, 654,  L143

\bibitem[{Sault}, {Teuben}, \&  {Wright} 1995]{1995ASPC...77..433S}
{Sault}, R.~J., {Teuben}, P.~J., \& {Wright}, M.~C.~H. 1995, in Astronomical  Society of the Pacific Conference Series, Vol.~77, Astronomical Data Analysis  Software and Systems IV, ed. {R.~A.~Shaw, H.~E.~Payne, \& J.~J.~E.~Hayes},  433

\bibitem[{Shepherd}, {Claussen}, \&  {Kurtz} 2001]{2001Sci...292.1513S}
{Shepherd}, D.~S., {Claussen}, M.~J., \& {Kurtz}, S.~E. 2001, Science, 292,  1513

\bibitem[{Stecklum}, {Feldt}, {Richichi}, {Calamai},  \& {Lagage} 1997]{1997ApJ...479..339S}
{Stecklum}, B., {Feldt}, M., {Richichi}, A., {Calamai}, G., \& {Lagage}, P.~O.  1997, \apj, 479, 339

\bibitem[{Thompson}, {Hatchell}, {Walsh},  {MacDonald}, \& {Millar} 2006]{2006A&A...453.1003T}
{Thompson}, M.~A., {Hatchell}, J., {Walsh}, A.~J., {MacDonald}, G.~H., \&  {Millar}, T.~J. 2006, \aap, 453, 1003

\bibitem[{Val'tts}, {Ellingsen}, {Slysh}, {Kalenskii},  {Otrupcek}, \& {Larionov} 2000]{2000MNRAS.317..315V}
{Val'tts}, I.~E., {Ellingsen}, S.~P., {Slysh}, V.~I., {Kalenskii}, S.~V.,  {Otrupcek}, R., \& {Larionov}, G.~M. 2000, \mnras, 317, 315

\bibitem[{van Diedenhoven}, {Peeters}, {Van  Kerckhoven}, {Hony}, {Hudgins}, {Allamandola}, \&  {Tielens} 2004]{2004ApJ...611..928V}
{van Diedenhoven}, B., {Peeters}, E., {Van Kerckhoven}, C., {Hony}, S.,  {Hudgins}, D.~M., {Allamandola}, L.~J., \& {Tielens}, A.~G.~G.~M. 2004, \apj,  611, 928

\bibitem[{Yamashita}, {Murata}, {Kawabe}, {Kaifu},  \& {Tamura} 1991]{1991ApJ...373..560Y}
{Yamashita}, T., {Murata}, Y., {Kawabe}, R., {Kaifu}, N., \& {Tamura}, M. 1991,  \apj, 373, 560

\bibitem[{Yamashita}, {Suzuki}, {Kaifu}, {Tamura},  {Mountain}, \& {Moore} 1989]{1989ApJ...347..894Y}
{Yamashita}, T., {Suzuki}, H., {Kaifu}, N., {Tamura}, M., {Mountain}, C.~M., \&  {Moore}, T.~J.~T. 1989, \apj, 347, 894

\bibitem[{Yorke} \& {Bodenheimer} 1999]{1999ApJ...525..330Y}
{Yorke}, H.~W. \& {Bodenheimer}, P. 1999, \apj, 525, 330

\bibitem[{Zapata}, {Ho}, {Schilke}, {Rodr{\'{\i}}guez},  {Menten}, {Palau}, \& {Garrod} 2009]{2009ApJ...698.1422Z}
{Zapata}, L.~A., {Ho}, P.~T.~P., {Schilke}, P., {Rodr{\'{\i}}guez}, L.~F.,  {Menten}, K., {Palau}, A., \& {Garrod}, R.~T. 2009, \apj, 698, 1422

\bibitem[{Zhang} 2005]{2005IAUS..227..135Z}
{Zhang}, Q. 2005, in IAU Symposium, Vol. 227, Massive Star Birth: A Crossroads  of Astrophysics, ed. {R.~Cesaroni, M.~Felli, E.~Churchwell, \& M.~Walmsley},  135--144

\bibitem[{Zinnecker} \& {Yorke} 2007]{2007ARA&A..45..481Z}
{Zinnecker}, H. \& {Yorke}, H.~W. 2007, \araa, 45, 481

\end{thebibliography}
\end{document}